\documentclass[aps,prx,reprint,amsmath,amssymb,longbibliography,superscriptaddress]{revtex4-2}
\usepackage{amsthm,booktabs,graphicx}
\usepackage[hidelinks]{hyperref}
\newtheorem{theorem}{Theorem}
\newtheorem{lemma}{Lemma}
\newcommand{\C}{\mathbb C}
\newcommand{\F}{\mathbb F}
\newcommand{\Z}{\mathbb Z}
\newcommand{\diag}{\operatorname{diag}}
\newcommand{\wt}{\operatorname{wt}}
\newcommand{\ket}[1]{\lvert#1\rangle}

\usepackage{xcolor}

\newcounter{algorithm}
\newenvironment{algorithm}[1][]{%
  \par\medskip\noindent\begin{minipage}{\linewidth}%
  \hrule height 0.8pt\kern4pt
  \renewcommand{\caption}[1]{%
    \refstepcounter{algorithm}%
    \textbf{Algorithm \thealgorithm.} ##1\par\kern4pt\hrule\kern4pt}%
}{\par\kern4pt\hrule height 0.8pt\end{minipage}\par\medskip}

\newtheorem{corollary}{Corollary}
\theoremstyle{definition}
\newtheorem{definition}{Definition}
\theoremstyle{plain}
\newcommand{\referenceedit}[1]{{#1}}
\newcommand{\rh}{\mathcal R_h}
\newcommand{\supp}{\operatorname{supp}}

\newcommand{\GL}{\operatorname{GL}}
\newcommand{\dig}{\operatorname{dig}}
\newcommand{\Tr}{\operatorname{Tr}}

\newcommand{\calU}{\mathcal{U}}

\makeatletter
\newcommand{\SMcontentsstyle}{%
  \normalfont\fontsize{11}{13.2}\selectfont
  \def\@pnumwidth{2em}%
  \def\@tocrmarg{3em}%
  \def\@dotsep{4.5}%
  \renewcommand*\numberline[1]{\hb@xt@\@tempdima{##1\hfil}}%
  \renewcommand*\l@section[2]{%
    \ifnum\c@tocdepth>\z@
      \addpenalty\@secpenalty
      \addvspace{1em \@plus\p@}%
      \setlength\@tempdima{2.3em}%
      \begingroup
        \parindent\z@ \rightskip\@pnumwidth
        \parfillskip-\@pnumwidth
        \leavevmode\bfseries
        \advance\leftskip\@tempdima
        \hskip-\leftskip
        ##1\nobreak\hfil
        \nobreak\hb@xt@\@pnumwidth{\hss##2}\par
      \endgroup
    \fi}%
  \renewcommand*\l@subsection{\@dottedtocline{2}{2.3em}{3em}}%
}
\makeatother

\begin{document}
\title{Substantially Improved Rate and Distance for Addressable Transversal Phase Gates}
\author{Bujiao Wu}
\email{wubujiao@iqasz.cn}
\affiliation{International Quantum Academy, Shenzhen, 518048, China}
\author{Tongyin Lin}
\email{12232812@mail.sustech.edu.cn}
\affiliation{International Quantum Academy, Shenzhen, 518048, China}
\author{Bin Cheng}
\email{bincheng@nus.edu.sg}
\affiliation{Centre for Quantum Technologies, National University of Singapore, Singapore 117543}
\begin{abstract}
An important ingredient in fault-tolerant quantum computation is
asymptotically good codes with fully addressable transversal
non-Clifford gates.
Existing constructions, however, provide only small explicit
guarantees for rate and relative distance.
Here, we substantially improve these guarantees for fully
addressable transversal phase gates at every fixed level
$h\ge3$ of the Clifford hierarchy, using a cyclic encoding
beyond the binary CSS framework.
The encoding separates logical phase control from the joint
design of rate and distance, enabling two constructions with
different computational costs.
A deterministic polynomial-time algebraic-geometric construction
gives a common asymptotic lower bound of $52/(675h)$ on both quantities.
A counting argument gives an exponential-time construction with
exact verification, improving this bound by a factor of approximately $2.86$.
We further establish a universal upper bound for codes supporting
fully addressable transversal single-qubit phase gates.
Within our cyclic encoding framework, we obtain a stronger upper
bound that is approximately $1.28$ times the lower bound achieved
by our exponential-time construction.
As concrete examples, we construct codes with $48$ and $90$
physical qubits supporting fully addressable transversal $T$ gates.
These results narrow the gap between achievable code parameters
and addressability bounds, providing quantitative targets for
further improvements in efficient code constructions.

\end{abstract}
\maketitle

\section{Introduction}
\label{sec:introduction}

Quantum error correction is essential for protecting quantum information against noise and enabling reliable and scalable quantum computation.
Of particular importance are transversal gates, which implement logical operations through physical gates that each act on at most one qubit per code block, preventing a single gate fault from spreading within a block~\cite{Daniel_Theory_1998}.
However, the Eastin-Knill theorem rules out a universal transversal gate set on a fixed code that detects arbitrary single-qubit errors~\cite{Bryan_Restrictions_2009}.
Nevertheless, transversal non-Clifford gates remain valuable for universal fault-tolerant computation through magic-state distillation~\cite{Sergey_Universal_2005,Sergey_Magic-state_2012}, code switching, and gauge fixing~\cite{Jonas_Fault-tolerant_2014,Hector_Gauge_2015}.
A further goal is to apply these logical gates selectively to individual logical qubits~\cite{Aranya_No-Go_2026}.
This property, known as addressability, allows a prescribed gate to act on any chosen logical qubit while leaving the others unchanged~\cite{Zhiyang_Quantum_2025,Tongyin_Asymptotically_2026}.
In addition, asymptotically good code families combine bounded data-qubit overhead per logical qubit with a minimum distance that grows linearly with the blocklength~\cite{AR_Good_1996,Andrew_Multiple_1996,Daniel_Fault-tolerant_2014}.
This motivates studying codes that support addressable transversal non-Clifford gates while maintaining a nonvanishing encoding rate and relative distance.

Algebraic-geometric (AG) codes already support combining positive asymptotic rate and relative distance with non-Clifford transversal operations.
He \emph{et al.} constructed asymptotically good codes with addressable transversal $\mathrm{CCZ}$ gates~\cite{Zhiyang_Asymptotically_2025}.
For single-qubit phase gates $Z(h)=\diag(1,e^{2\pi i/2^h})$ in the Clifford hierarchy, San-Jos\'e obtained asymptotically
good CSS codes with addressable level-$h$ rotations for every
fixed $h\ge3$ by lowering a level-$(h+1)$
construction~\cite{Rodrigo_Asymptotically_2026}, whereas Lin,
Wu, and Cheng achieved fully addressable transversal $T$ gates
on asymptotically good CSS codes directly through generalized
divisibility~\cite{Tongyin_Asymptotically_2026}.
Meiburg gave both CSS constructions with fully addressable transversal $T$ and a modular-clock construction for single-qubit gates of arbitrary fixed finite projective order~\cite{Alexander_Constant_2026}.
However, as shown in Table~\ref{tab:prior}, although these constructions give asymptotically
good codes, the simultaneous rate and relative-distance
guarantees remain unrealistically small, limiting their
practical applicability. This motivates a natural question: \textit{Can these parameters be improved to practically
meaningful values, such as a few percent, while supporting
fully addressable transversal phase gates?}

\begin{table}[t]
\centering
\setlength{\tabcolsep}{4pt}
\renewcommand{\arraystretch}{1.1}
\caption{Common asymptotic lower bounds on encoding rate and
relative distance for codes supporting fully addressable
transversal $T$ gates implemented by physical $T$ powers.
Values are rounded; parameter calculations are given in
Sec.~S5 of the Supplemental Material (SM).}
\label{tab:prior}
\begin{tabular*}{\columnwidth}{@{\extracolsep{\fill}}lr@{}}
\toprule
Construction & Common lower bound\\
\midrule
San-Jos\'e~\cite{Rodrigo_Asymptotically_2026}
 & $4.6027\times10^{-9}$\\
Lin et al.~\cite{Tongyin_Asymptotically_2026}
 & $4.1356\times10^{-6}$\\
Meiburg (CSS)~\cite{Alexander_Constant_2026}
 & $7.1751\times10^{-5}$\\
Meiburg (modular clock)~\cite{Alexander_Constant_2026}
 & $9.6006\times10^{-4}$\\
This work (polynomial)
 & $2.5679\times10^{-2}$\\
This work (exponential)
 & $7.3352\times10^{-2}$\\
\bottomrule
\end{tabular*}
\end{table}

\begin{figure*}[t]
\centering
\includegraphics[width=\textwidth]{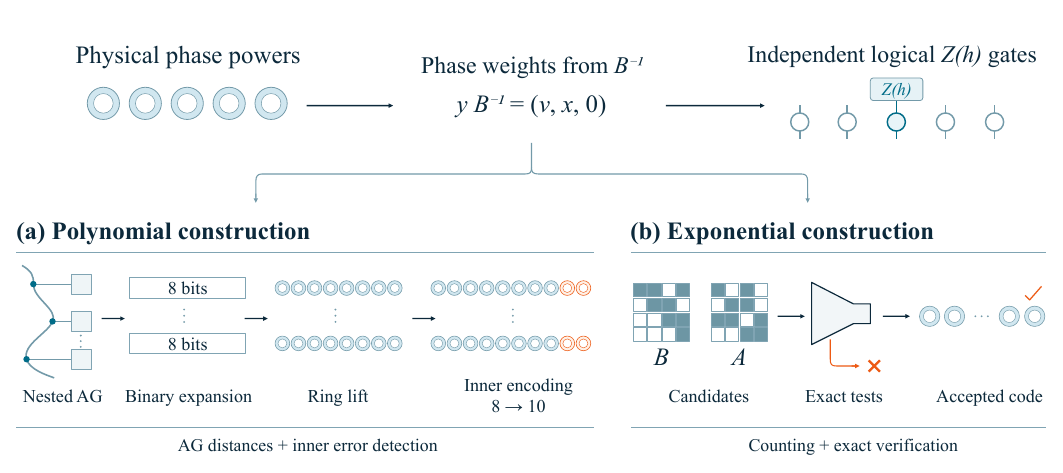}
\caption{Overview of quantum code constructions with fully addressable
transversal $Z(h)$ gates.
Columns of $\mathsf B^{-1}$ determine physical phase powers for
independently prescribed logical $Z(h)$ operations; at fixed $\mathsf B$
these powers are independent of $A$.
(a) Nested AG codes over $\F_{256}$ undergo binary expansion and an
adapted-basis ring lift with $A=0$. The $8$-to-$10$ inner encoding adds
the orange registers and preserves the logical phase action.
(b) Counting guarantees a pair $(\mathsf B,A)$ passing the exact Weyl
rejection tests at the stated parameters.
 Each ring symbol represents a register of $h$ qubits.}
\label{fig:construction-overview}
\end{figure*}
In this work, we construct quantum codes with fully addressable transversal phase gates and substantially improved rate and relative distance.
Our approach uses an encoding over the cyclic ring $\mathcal R_h=\Z/2^h\Z$
beyond the binary CSS framework, representing each ring coordinate by
$h$ physical qubits. Columns of the inverse encoding matrix recover the
binary logical labels as linear functions of these coordinates.
Expanding the coordinates into binary digits turns these functions into
physical single-qubit phase gates with independently prescribed logical
actions. This separates addressability from the joint design of rate
and distance, replacing the higher-order overlap and divisibility
conditions of the previous CSS constructions with a linear coordinate
identity.

For the polynomial-time construction, we start with nested AG codes
over $\F_{256}$~\cite{Arnaldo_On_1996,Maria_The_2006,Kenneth_A_2001}.
A binary basis expansion gives blocks of eight bits. Completing a basis
adapted to the nested codes and lifting it to $\mathcal R_h$ yields an invertible
encoding matrix while retaining the relevant primal and dual block
distances. Uniform superpositions over the resulting ring-code cosets
define the outer quantum code.
We then encode each eight-register block into ten registers.
A coherent uniform offset and a linear check detect single-register
errors, while a compensating phase on the offset register preserves
the prescribed logical action. This doubles the certified
physical-qubit distance lower bound at a $25\%$ increase in blocklength,
with the number of logical qubits unchanged.

For the $T$ gate, a deterministic polynomial-time construction gives a common asymptotic lower bound of approximately $2.57\%$ on both quantities.
A counting argument over encoding matrices and quadratic phases gives
an exponential-time construction with exact verification, increasing
this guarantee to approximately $7.34\%$.
For every fixed hierarchy level $h\ge3$, both constructions
give common rate and relative-distance guarantees that scale
as $1/h$.
This mild dependence on $h$ is particularly favorable at higher
hierarchy levels, where the explicit guarantees of previous
constructions deteriorate rapidly~\cite{Rodrigo_Asymptotically_2026,
Alexander_Constant_2026}.
We also derive a universal upper bound on the common asymptotic
rate--distance coefficient for codes supporting fully addressable
transversal $Z(h)$ powers.
The bound follows by reducing the addressing matrix modulo two and
applying classical rate--distance bounds to the resulting binary code.
Within our cyclic encoding framework, the stronger upper bound
scales as $1/h$ and is approximately $1.28$ times the lower bound
achieved by our exponential-time construction.
In addition, we construct codes with $48$ and $90$ physical qubits supporting fully addressable transversal $T$ gates as well as their encoding circuits, providing concrete examples of our framework.

The remainder of the paper is organized as follows.
Section~\ref{sec:framework} introduces the encoding and its
classical ingredients. Section~\ref{sec:constructions} proves
the polynomial-time construction and the existence result with exact
exponential-time generation.
Section~\ref{sec:bounds} derives upper bounds on the simultaneously achievable rate and relative distance, and
Sec.~\ref{sec:finite} illustrates the framework with two finite
codes. Technical lemmas, construction algorithms, counting details,
and exact certification criteria appear in SM.

\section{Preliminaries and cyclic encoding}
\label{sec:framework}
Here we introduce the code parameters and addressability condition,
review the required properties of AG codes, and present the
cyclic encoding together with its transversal logical phase action.
\subsection{Code parameters and addressability}
A binary quantum code $((N,2^K,D))_2$ encodes $K$ logical qubits
into $N$ physical qubits and detects every operator supported on
fewer than $D$ qubits. Let $E$ be an encoding isometry for the code.
The code detects an operator
$O$ if $E^\dagger OE = \lambda_O I_{2^K}$ for some $\lambda_O \in \C$~\cite{Emanuel_Theory_1997}.
The code's rate and
relative distance are
$\mathsf R=K/N$ and $\mathsf d=D/N$. For a code family with $N_i\to\infty$,
let
\begin{align*}
 \mathsf R_\infty:=\liminf_i K_i/N_i,\qquad
 \mathsf d_\infty:=\liminf_i D_i/N_i,
\end{align*}
and define $\eta:=\min\{\mathsf R_\infty,\mathsf d_\infty\}$.
In this work, all logarithms are base two,
and each asymptotic statement fixes $h$.

For an integer $h\ge1$, set
\[
 \nu:=2^h,\qquad \mathcal R_h:=\Z/\nu\Z,\qquad
 Z(h):=\diag(1,e^{2\pi i/\nu}).
\]
The levels $h=1,2,3$ give the gates $Z,S$ and $T$.
Full addressability means that
there exist one encoding isometry $E$ and one matrix
$F\in\mathcal R_h^{K\times N}$ satisfying
\begin{equation}
 \left(\bigotimes_{j=1}^N Z(h)^{(cF)_j}\right)E
 =E\left(\bigotimes_{\ell=1}^K Z(h)^{c_\ell}\right)
 \quad(c\in\mathcal R_h^K).
 \label{eq:address}
\end{equation}
Thus, the vector $c$ independently chooses each logical phase
power, with zero selecting an idle qubit. This identity gives
the desired logical action exactly, on a fixed encoding for
the chosen level $h$.

\subsection{Nested algebraic-geometric codes}
We use nested one-point evaluation codes. Let $X/\F_q$ be a
smooth projective curve of genus $g$, let $P_\infty$ be a rational
point, and let $\mathcal D$ be a sum of $n$ other distinct rational
points. The Riemann--Roch space $\mathcal{L}(aP_\infty)$ consists
of functions whose only allowed pole is at $P_\infty$, of order
at most $a$. Evaluating these functions at the points in the support of
$\mathcal D$ gives the algebraic-geometric (AG) code $C_{\mathrm{L}}(\mathcal D,aP_\infty)$.
The following standard consequence of Riemann--Roch gives the
parameters needed below~\cite{Henning_Algebraic_2009}.
\begin{lemma}[Ref. \cite{Henning_Algebraic_2009}]
\label{lem:ag-input}
If $2g-2<a<n$, then $C=C_{\mathrm{L}}(\mathcal D,aP_\infty)$ satisfies $ \dim C=a+1-g$, and the {distances} of code $C$ and $C^{\perp}$ are at least $n - a$ and $a + 2 - 2g$.
Moreover, $a_0<a_1$ gives
$C_{\mathrm{L}}(\mathcal D,a_0P_\infty)\subseteq C_{\mathrm{L}}(\mathcal D,a_1P_\infty)$.
\end{lemma}
The Garcia--Stichtenoth tower over $\F_{256}$ provides a sequence
of curves with $n_i\to\infty$ and $g_i/n_i\to1/15$~\cite{Arnaldo_On_1996,Maria_The_2006}.
With suitable choices of divisor degree, Lemma~\ref{lem:ag-input}
therefore yields AG code families with positive asymptotic rate
and relative minimum distance.
Its evaluation matrices can be constructed deterministically
in polynomial time~\cite{Kenneth_A_2001}.
Thus, these AG code families are asymptotically good and can be constructed deterministically and efficiently.

\subsection{Encoding and logical addressability}
Each element $y_j\in\mathcal R_h$ has a binary expansion
$y_j=\sum_{b=0}^{h-1}2^by_{j,b}$. We write
$\operatorname{dig}(y_j)=(y_{j,0},\ldots,y_{j,h-1})$.
Each ring coordinate is represented by a register of $h$
physical qubits. Hence, $M$ ring coordinates occupy $N=hM$
physical qubits. Arithmetic on ring coordinates is modulo $2^h$.

Choose an invertible $M\times M$ matrix $\mathsf B$ over $\mathcal R_h$.
Let $G$ be its first $r$ rows and $L$ its next $K$ rows, with
$r+K\le M$.
Let $\mathcal R_h^r:=\{(v_1,\ldots,v_r):v_j\in\mathcal R_h\text{ for }1\le j\le r\}$.
Given a symmetric $M\times M$ ring matrix $A$, define a map $E:\C^{2^K}\to\C^{2^N}$ such that
\begin{equation}
 E\ket{x}=\nu^{-r/2}\sum_{v\in\mathcal R_h^r}
 e^{\pi i yAy^{\mathsf T}/\nu}\ket{\operatorname{dig}(y)},
 \label{eq:master}
\end{equation}
for every $x\in\{0,1\}^K$ and $y=vG+xL \in \mathcal{R}_h^M$, where in the quadratic phase, we use integer representatives in
$\{0,\ldots,\nu-1\}$ for all entries of $A$ and $y$.
The AG construction uses $A=0$ in Sec.~\ref{sec:constructions}, while the exponential-time
construction searches over symmetric $A$ and invertible
$\mathsf B$ with exact verification of the required distance bound.
For each $1 \leq \ell \leq K$, let $z_\ell^{\mathsf T}$ be column $r+\ell$
of $\mathsf B^{-1}$.

\begin{lemma}
\label{lem:selector}
Equation~\eqref{eq:master} defines an encoding isometry $E$.
The code defined by $E$ admits fully addressable transversal $Z(h)$ gates,
with addressing matrix $F\in\mathcal R_h^{K\times hM}$ given by
\begin{equation}
 F_{\ell,(j,b)}=2^bz_{\ell,j}\pmod{\nu},
 \label{eq:digitform}
\end{equation}
where $1\le j\le M$ and $0\le b<h$.
\end{lemma}
\begin{proof}
Invertibility of $\mathsf B$ makes the map $(v,x)\mapsto vG+xL$
injective. The encoded states in Eq.~\eqref{eq:master} corresponding to distinct
binary labels $x$ therefore have disjoint supports, and each state contains
$\nu^r$ terms of magnitude $\nu^{-r/2}$. The encoded states are thus orthonormal, so $E$
is an isometry.

Since $y=(v,x,0)\mathsf B$ and $z_\ell^{\mathsf T}$ is column
$r+\ell$ of $\mathsf B^{-1}$, we have
\[
 \sum_{j=1}^{M}z_{\ell,j}y_j
 =yz_\ell^{\mathsf T}\equiv x_\ell\pmod{\nu}.
\]
Substituting $y_j=\sum_{b=0}^{h-1}2^by_{j,b}$ gives
\[
 x_\ell\equiv\sum_{j=1}^{M}\sum_{b=0}^{h-1}
 (2^bz_{\ell,j})y_{j,b}\pmod{\nu}.
\]
The coefficient of each physical bit therefore gives the
addressing entry $F_{\ell,(j,b)}=2^bz_{\ell,j}\pmod{\nu}$.
Indeed, applying $Z(h)^{F_{\ell,(j,b)}}$ to physical qubit
$(j,b)$ contributes the phase
$e^{2\pi iF_{\ell,(j,b)}y_{j,b}/\nu}$. Their product is
$e^{2\pi ix_\ell/\nu}$ on every term of $E\ket{x}$, implementing
$Z(h)$ on logical qubit $\ell$ and the identity on the others.
Combining the rows with coefficients $c_\ell$ gives
Eq.~\eqref{eq:address}.
\end{proof}

The vectors $z_\ell$, and hence the addressing matrix $F$,
depend only on $\mathsf B^{-1}$. For fixed $\mathsf B$, the
same $F$ therefore implements the prescribed logical phase
operations for every symmetric matrix $A$. The choice of $A$
changes the relative phases within the encoded states,
providing additional freedom to obtain distance guarantees
while preserving logical addressability.

Once $\mathsf B$ and $A$ are specified, the encoding admits
a polynomial-size circuit. Indeed, for fixed $h$, modular elimination
implements the linear map $w\mapsto w\mathsf B$ using
$O_h(M^2)$ elementary gates~\cite[Sec.~IV]{Erik_Stabilizer_2005}\footnote{The subscript $h$ indicates that the hidden constant may depend on the fixed phase level $h$.}.
The uniform superposition requires $hr=O_h(M)$ Hadamard gates, and binary expansion decomposes the quadratic phase in Eq.~\eqref{eq:master} into $O_h(M^2)$ diagonal gates acting on one or two qubits.

\section{Asymptotically Good Code Constructions}
\label{sec:constructions}
This section gives two constructions of quantum codes with fully addressable $Z(h)$
gates as well as positive asymptotic rate and relative distance for each
fixed $h\ge3$. The first combines nested AG codes with an inner code
and can be implemented deterministically in polynomial time (Theorem~\ref{thm:poly}).
The second uses random encoding
matrices and quadratic phases to obtain a larger common lower bound
on rate and relative distance, with exact exponential-time generation
(Theorem~\ref{thm:exist}). Figure~\ref{fig:construction-overview}
shows the shared phase mechanism and two approaches to obtaining distance.

\subsection{Polynomial-time algebraic-geometric construction}
Our construction concatenates an outer quantum code with an
inner code. The outer code is constructed from nested one-point AG codes
over $\F_{256}$, with each field coordinate represented by
eight ring registers, as detailed in the proof of Theorem~\ref{thm:poly}
and in Sec.~S2.1 of SM.

The inner encoding acts on each eight-register block by
introducing a uniformly superposed common offset and two
additional registers that store the offset and a modular
linear check. The resulting ten-register code detects every
error supported on a single output register and allows the
outer logical phase operations to remain transversal.
Each ring register is represented by $h$ physical qubits; we write
$\ket{s}$ as shorthand for $\ket{\operatorname{dig}(s)}$, where $s\in\mathcal R_h$.

We now define this inner encoding and prove these two properties.
Its extension to arbitrary even block sizes and the full
concatenation argument appear in
Sec.~S2.2 of SM.

Set $\epsilon_a=1$ for $1\le a\le4$ and $\epsilon_a=-1$ for
$5\le a\le8$, so $\sum_a\epsilon_a=0$. For $u\in\mathcal R_h^8$,
define the inner encoding as
\begin{equation}
 \begin{aligned}
 J_8\ket u=\nu^{-1/2}\sum_{t\in\mathcal R_h}
 &\ket{u_1+t,\ldots,u_8+t}\\[-2pt]
 &{}\otimes\ket t
 \otimes\ket{t-\sum_{a=1}^8\epsilon_a u_a}.
 \end{aligned}
 \label{eq:ag-inner}
\end{equation}
The inner code defined by $J_8$ detects errors supported on any
single output register, as the following lemma shows.

\begin{lemma}
\label{lem:inner-detection}
The map $J_8$ is an isometry, and the code it defines detects
every operator supported on a single output register.
\end{lemma}
\begin{proof}
The ninth output register takes the value $t$, and subtracting $t$
from the first eight registers recovers $u$. Distinct inputs
therefore have disjoint supports,
and each encoded state contains $\nu$ distinct terms of amplitude $\nu^{-1/2}$, proving isometry.
Every output word $w$ satisfies
\[
 \sum_{a=1}^8\epsilon_aw_a-w_9+w_{10}=0.
\]
The single-register Weyl operators $Q_{a,b}$, defined by
$Q_{a,b}\ket s=e^{2\pi i bs/\nu}\ket{s+a}$ for
$a,b\in\mathcal R_h$, combine a shift by $a$ with a phase
factor $e^{2\pi i bs/\nu}$.
They form a basis for all operators on one register~\cite{Erik_Stabilizer_2005}.
{If $a\ne0$, the operator shifts one coordinate of every output word by $a$.
For every shifted word, the left side of the above identity equals $\pm a\ne0$.
All basis words in all encoded states satisfy the identity,
so the code space is mapped into its orthogonal complement.}
If $a=0$ and $b\ne0$, write the affected output coordinate as
$t+f(u)$. The phase operator $Q$ multiplies each term by
$e^{2\pi i b(t+f(u))/\nu}$. Because the superposition is uniform,
the inner product with the original encoded state is the average
of these phase factors, and it vanishes:
\[
 \begin{aligned}
 \langle u|J_8^\dagger QJ_8|u\rangle
 &=\nu^{-1} e^{2\pi i b f(u)/\nu}
   \sum_{t\in\mathcal R_h}e^{2\pi i bt/\nu}\\
 &=0.
 \end{aligned}
\]
Inner products with other encoded basis states also vanish,
since phase operators preserve the disjoint supports of distinct inputs.
Thus $J_8^\dagger QJ_8=0$ for every nonidentity Weyl operator $Q$
on one register. By linearity, $J_8^\dagger OJ_8$ is a scalar
multiple of the identity for every single-register operator $O$,
which proves detection.
\end{proof}

For a ring register, write
$Z_\nu({b})\ket s=e^{2\pi i {b}s/\nu}\ket s$; in binary digits,
$Z_\nu({b})=\bigotimes_{{j}=0}^{h-1}Z(h)^{2^{{j}}{b}}$.
The next identity transfers a product of these phase operators
through the inner encoding.
\begin{lemma}
\label{lem:inner-selector}
For $z\in\mathcal R_h^8$, set
$\widetilde z=(z_1,\ldots,z_8,-{\sum_{a=1}^8} z_a,0)$. Then
\begin{equation}
 \left(\bigotimes_{j=1}^{10}Z_\nu(\widetilde z_j)\right)J_8
 =J_8\left(\bigotimes_{a=1}^{8}Z_\nu(z_a)\right).
 \label{eq:ag-inner-selector}
\end{equation}
\end{lemma}
\begin{proof}
For each summand of Eq.~\eqref{eq:ag-inner}, the phase exponent is
\[
 \begin{aligned}
 \widetilde zw^{\mathsf T}
 &=\sum_{a=1}^8z_a(u_a+t)-\left(\sum_{a=1}^8z_a\right)t\\
 &=\sum_{a=1}^8z_au_a\pmod\nu.
 \end{aligned}
\]
It is independent of $t$ and equals the phase exponent on the
input state, proving the identity.
\end{proof}

Combining these properties with the AG distance bounds gives
the following family.
\begin{theorem}
\label{thm:poly}
For each fixed $h\ge3$, a deterministic polynomial-time
algorithm constructs fully addressable $Z(h)$ codes with
\begin{equation}
 \lim_i\frac{K_i}{N_i}=\frac{52}{675h},\qquad
 \liminf_i\frac{D_i}{N_i}\ge\frac{52}{675h}.
 \label{eq:poly}
\end{equation}
\end{theorem}
\begin{proof}
Fix $h\ge3$. We first construct the outer encoding, then use
$J_8$ to obtain a binary code with the required distance and
addressability, and finally choose the AG parameters.
On a curve over $\F_{256}$ with $n$ evaluation points and genus
$g$, choose $2g-2<a_0<a_1<n$ and set
$\mathcal A_b=C_{\mathrm{L}}(\mathcal D,a_bP_\infty)$ for $b=0,1$.
Lemma~\ref{lem:ag-input} gives nested codes with
\[
 \begin{gathered}
 \dim\mathcal A_b=a_b+1-g,\\
 d(\mathcal A_1)\ge n-a_1,\qquad
 d(\mathcal A_0^\perp)\ge a_0+2-2g.
 \end{gathered}
\]
The dimension difference determines the number of logical
qubits, and the two distance bounds control shifts and phases.

Expand each field coordinate in a fixed binary basis to obtain
$B_0\subset B_1\subseteq\F_2^{8n}$. Group the eight coordinates
of each field symbol into one block. Extend a basis of
$B_0$ to a basis of $B_1$ and then to a basis of
$\F_2^{8n}$. Arrange these basis vectors as rows of a matrix
and lift its zero--one entries to $\mathcal R_h$.
The resulting matrix $\mathsf B_{\mathrm{out}}$ has odd
determinant and is therefore invertible.
Let $G$ contain its first $r_{\mathrm{out}}$ rows and $L$ its
next $K$ rows, where
\[
 r_{\mathrm{out}}=8(a_0+1-g),\qquad K=8(a_1-a_0).
\]
Setting $A=0$ in Eq.~\eqref{eq:master} defines the outer encoding
\begin{equation}
 E_{\mathrm{out}}\ket x
 =\nu^{-r_{\mathrm{out}}/2}
 \sum_{v\in\mathcal R_h^{r_{\mathrm{out}}}}\ket{vG+xL},
 \qquad x\in\{0,1\}^K.
 \label{eq:ag-outer}
\end{equation}
By Lemma~\ref{lem:selector}, it is an isometry with fully
addressable logical $Z(h)$ gates.

Binary expansion preserves nonzero field-coordinate blocks,
and expanding $\mathcal A_0^\perp$ in the trace-dual basis
gives $B_0^\perp$~\cite{Rudolf_Introduction_1994}.
Thus every nonzero word of $B_1$ has at least $n-a_1$
nonzero blocks, while every nonzero word of $B_0^\perp$
has at least $a_0+2-2g$ nonzero blocks.

For any nonzero ring codeword, write
$c=2^k\alpha\binom{G}{L}$ with $0\le k<h$ and
$\alpha\bmod2\ne0$. Row independence modulo two ensures that
$b=\alpha\binom{G}{L}\bmod2$ is a nonzero word of $B_1$.
Whenever $b_j=1$, the coordinate $c_j$ is $2^k$ times
an odd element and hence nonzero modulo $2^h$.
Thus every nonzero block of $b$ remains nonzero in $c$,
giving at least $n-a_1$ nonzero blocks.

Similarly, for a nonzero ring vector $z$ satisfying
$Gz^{\mathsf T}=0$, write $z=2^kz'$ with $0\le k<h$
and at least one odd entry in $z'$. Then
$Gz'^{\mathsf T}\equiv0\pmod{2^{h-k}}$, so
$z'\bmod2$ is a nonzero word of $B_0^\perp$ whose support
is contained in that of $z$. Thus $z$ has at least
$a_0+2-2g$ nonzero blocks.
The full distance-preservation result is proved in
Sec.~S2.1 of SM.

Set
\[
 d_*=\min\{n-a_1,a_0+2-2g\}.
\]
A nonzero shift on fewer than $d_*$ blocks cannot connect outer-code basis states.
A nonidentity pure phase
error on those blocks has the form
$Q_\beta=\bigotimes_{j=1}^{8n}Z_\nu(\beta_j)$, with
$0\ne\beta\in\mathcal R_h^{8n}$.
It leaves each basis label unchanged and multiplies its amplitude by
$e^{2\pi i(vG+xL)\beta^{\mathsf T}/\nu}$.
The dual distance bound implies $G\beta^{\mathsf T}\ne0$,
so averaging this phase factor over all $v$ in $\mathcal R_h^{r_{\mathrm{out}}}$ gives zero.
Consequently, $E_{\mathrm{out}}^\dagger QE_{\mathrm{out}}=0$
for every nonidentity Weyl operator $Q$ supported on fewer
than $d_*$ blocks. Expanding any operator $O$ on those blocks
in the Weyl basis, only its identity term contributes to
$E_{\mathrm{out}}^\dagger OE_{\mathrm{out}}$, so the result is a scalar
multiple of the logical identity. Thus the outer code detects every error
supported on fewer than $d_*$ blocks of eight ring registers.

Next apply $J_8$ to each of the $n$ outer blocks, and represent
each output register by its $h$ binary digits. This gives an
encoding $E$ into $N=10hn$ physical qubits, with the same $K$
logical qubits.
To prove the distance bound, consider any physical Pauli operator $P$
of weight $w<2d_*$. Write $P=\bigotimes_{p=1}^nP_p$, where $P_p$
acts on the $10h$ physical qubits in block $p$.
If $P_p$ acts nontrivially on at most one qubit, its support lies
within one output ring register. Lemma~\ref{lem:inner-detection}
therefore implies that $J_8^\dagger P_pJ_8$ is a scalar multiple
of the identity. Thus only blocks containing
at least two affected qubits can contribute nonscalar factors.
There are at most $\lfloor w/2\rfloor<d_*$ such blocks.
The outer detection result therefore
implies that $E^\dagger PE$ is a scalar multiple of the logical identity.
By linearity, every Pauli error on fewer than
$2d_*$ physical qubits is detectable. Hence
\begin{equation}
 \begin{gathered}
 N=10hn,\qquad K=8(a_1-a_0),\\
 D\ge2\min\{n-a_1,a_0+2-2g\}.
 \end{gathered}
 \label{eq:agfinite}
\end{equation}

To retain addressability after concatenation, let $z_\ell^{\mathsf T}$
be column $r_{\mathrm{out}}+\ell$ of $\mathsf B_{\mathrm{out}}^{-1}$
and partition $z_\ell$ into eight-coordinate blocks $z_{\ell,p}$.
Extend each block to $\widetilde z_{\ell,p}$ using
Lemma~\ref{lem:inner-selector}. By Eq.~\eqref{eq:ag-inner-selector},
the corresponding phase operators implement the same outer logical action
after concatenation. Expanding each output register into binary digits
gives the physical addressing matrix $F$ with entries
\[
 \begin{gathered}
 F_{\ell,(p,j,b)}=2^b\widetilde z_{\ell,p,j}\pmod\nu,\\
 1\le p\le n,\quad 1\le j\le10,\quad 0\le b<h.
 \end{gathered}
\]
For each logical qubit $\ell$, this row implements $Z(h)$ on
that qubit and the identity on the others. Taking linear
combinations of the rows proves Eq.~\eqref{eq:address} for
every logical exponent vector on this fixed encoding.

Finally, take the Garcia--Stichtenoth tower over $\F_{256}$,
with $n_i\to\infty$ and $g_i/n_i\to1/15$~\cite{Arnaldo_On_1996,Maria_The_2006}.
For degree fractions $a_{b,i}/n_i\to\beta_b$ satisfying
$2/15<\beta_0<\beta_1<1$, Eq.~\eqref{eq:agfinite} gives
\begin{equation}
 \begin{aligned}
 \mathsf R_\infty&=\frac4{5h}(\beta_1-\beta_0),\\
 \mathsf d_\infty&\ge\frac1{5h}
       \min\{1-\beta_1,\beta_0-2/15\}.
 \end{aligned}
 \label{eq:agfractions}
\end{equation}
Balancing the rate and the two distance terms gives
\[
 4(\beta_1-\beta_0)=1-\beta_1=\beta_0-2/15,
\]
so choose $\beta_0=70/135$ and $\beta_1=83/135$, and set
\[
 a_{0,i}=\lfloor70n_i/135\rfloor,\qquad
 a_{1,i}=\lfloor83n_i/135\rfloor.
\]
These degrees satisfy $2g_i-2<a_{0,i}<a_{1,i}<n_i$ for all
sufficiently large $i$. Rounding changes each normalized degree
by less than $1/n_i$, so Eq.~\eqref{eq:agfractions} yields
Eq.~\eqref{eq:poly}.
\end{proof}

For fixed $h$, the evaluation matrices are constructed
deterministically in polynomial time~\cite{Kenneth_A_2001}.
Binary expansion, basis completion, ring-matrix inversion and
the inner encodings also take polynomial time in $N_i$.
The comparison of fixed square binary fields and the explicit
construction steps are given in Secs.~S2.3 and S2.4 of SM, respectively.

The same proof gives a rate--distance tradeoff. For
$0<\tau_{\mathrm d}<13/30$, choose
$\beta_0=2/15+\tau_{\mathrm d}$ and $\beta_1=1-\tau_{\mathrm d}$
in Eq.~\eqref{eq:agfractions}. Then
\begin{equation}
 \mathsf R_\infty=\frac4{5h}\left(\frac{13}{15}-2\tau_{\mathrm d}\right),
 \qquad \mathsf d_\infty\ge\frac{\tau_{\mathrm d}}{5h}.
 \label{eq:agregion}
\end{equation}
Increasing $\tau_{\mathrm d}$ improves the lower bound on the relative distance while reducing the rate.
The balanced choice is
$\tau_{\mathrm d}=52/135$; for $T=Z(3)$ it gives rate $52/2025$,
corresponding asymptotically to $2025/52\simeq38.94$ physical
data qubits per logical qubit.

\begin{table}[t]
\centering
\setlength{\tabcolsep}{6pt}
\caption{Values of $\eta$ achieved by the two constructions and
theoretical upper bounds for codes admitting fully addressable logical $Z(h)$ gates at any fixed $h\ge3$.
Here $\eta=\min\{\mathsf R_\infty,\mathsf d_\infty\}$ is the minimum
of the asymptotic rate and relative distance. Numerical values are rounded to the displayed precision.
}
\label{tab:asymptotic}
\begin{tabular}{@{}p{0.25\textwidth}cc@{}}
\toprule
Result and scope & $Z(h)$ & $T$ ($h=3$)\\
\midrule
Polynomial-time AG
& $52/(675h)$ & $0.0257$\\
Exponential-time
& $t_\star/h$ & $0.0734$\\
 General upper bound
& $0.2827$ & $0.2827$\\
Upper bound (cyclic)
& $0.2827/h$ & $0.0942$\\
\bottomrule
\end{tabular}
\end{table}

\subsection{Existence with exact exponential-time generation}
\label{sec:existence}
The second construction uses a random quadratic phase to obtain codes that detect
low-weight errors. Fix $h\ge3$, set $\nu=2^h$, and take
$1\le K<M$, $1\le d\le M$, and $r=M-K$.
Choose $\mathsf B$ uniformly among invertible $M\times M$ matrices
over $\mathcal R_h$, and choose $A$ independently and uniformly
among symmetric matrices. Write
$\mathsf B=\begin{pmatrix}G\\L\end{pmatrix}$, with $G$ having $r$ rows.
The encoding $E$ is given by Eq.~\eqref{eq:master}, with $N=hM$.

For $a,b\in\mathcal R_h^M$, the Weyl operators
$W(a,b)=\bigotimes_{j=1}^M X_\nu(a_j)Z_\nu(b_j)$ form an
orthogonal basis for all errors~\cite{Erik_Stabilizer_2005}, where $X_\nu(a_j)\ket{ y_j}=\ket{y_j+a_j}$ and $Z_\nu(b_j)\ket{y_j}=e^{2\pi ib_jy_j/\nu}\ket{y_j}$ for $y_j \in \mathcal{R}_h$.
Every operator $P$ therefore has the Weyl expansion
\[
 P=2^{-N}\sum_{a,b}
 \operatorname{Tr}\!\left(W(a,b)^\dagger P\right)W(a,b).
\]

We first characterize when a Weyl operator has a nonzero matrix
element between encoded basis states.

\begin{lemma}
\label{lem:weyl-compression}
For $x,x'\in\{0,1\}^K$, the matrix element
$\langle x'|E^\dagger W(a,b)E|x\rangle$ is nonzero if and only if
\begin{equation}
 G(b-aA)^{\mathsf T}=0,\qquad
 (a\mathsf B^{-1})_{r+1:M}=x'-x
 \label{eq:compression-main}
\end{equation}
over $\mathcal R_h$.
\end{lemma}
The proof is given in Sec.~S3.1 of SM.

To identify the Weyl terms that must be tested, we assign
each pair a cost that lower-bounds the weight of any qubit
Pauli whose Weyl expansion contains that pair.
Thus all terms arising from Pauli errors of weight below $d$
are included among the pairs of cost below $d$.
We choose the cost according to the additive order of the pair.
{%
A pair $(a,b)$ has exact order $2^s$ if
$2^s(a,b)=0$ and $2^{s-1}(a,b)\ne0$ over $\mathcal R_h$.
We assign each pair a total cost:
\begin{enumerate}
    \item [(1)] For exact order two, write
    $(a,b)=2^{h-1}(\alpha,\beta)$ with
    $\alpha,\beta\in\{0,1\}^M$.
    The total cost is $\wt(\alpha)+\wt(\beta)$.

    \item[(2)] For exact order four, each local pair satisfies
    $4(a_j,b_j)=0$.
    Assign local cost zero to $(0,0)$.
    Assign local cost one to a nonzero local pair if its Weyl
    operator appears with a nonzero coefficient in the Weyl
    expansion of a single-qubit Pauli on that register.
    Assign local cost two to all remaining nonzero local pairs.
    The total cost is the sum of these local costs.

    \item[(3)] For exact order $2^s$ with $3\le s\le h$, the total
    cost is the number of registers with $(a_j,b_j)\ne(0,0)$.
\end{enumerate}

For a pair of exact order two, every entry of the logical shift
$(a\mathsf B^{-1})_{r+1:M}$ is zero or $2^{h-1}$.
Since $h\ge3$, this vector can equal a difference of binary
labels $x'-x$ only if it is zero, in which case $x'=x$.

We now define a sufficient test for the distance bound $D\ge d$.
Among nonzero Weyl pairs of total cost below $d$, an
exact-order-two pair triggers rejection if both
$G(b-aA)^{\mathsf T}=0$ and
$(a\mathsf B^{-1})_{r+1:M}=0$ hold.
A pair of higher order triggers rejection if
$G(b-aA)^{\mathsf T}=0$.
Let $\mathsf{reject}$ denote the event that at least one such
pair triggers rejection.
When this event occurs, we discard the candidate matrix pair
$(\mathsf B,A)$.

}

Passing this test guarantees $D\ge d$, as shown in the proof
of Theorem~\ref{thm:exist}.
To establish existence, we bound
$\Pr(\mathsf{reject})$ using the following counting functions
and estimates for each exact order:
\begin{align}
\begin{split}
      B(n,d)&=\sum_{w=1}^{d-1}\binom nw, \\
      W_h(M,d)&=\sum_{w=1}^{d-1}[z^w] (1+c_hz+(15-c_h)z^2)^M, \\
      V_s(M,d)&=\sum_{w=1}^{d-1}(4^s-1)^w\binom Mw,
\end{split}
\label{eq:weyl-volumes-main}
\end{align}
where $c_3=10$ and $c_h=6$ for $h\ge4$.
\begin{lemma}[Order-two errors]
\label{lem:order-two}
Among exact-order-two pairs of cost below $d$, the
$\alpha\ne0$ branch contains at most $B(2M,d)$ pairs,
each with rejection probability at most $2^{-M}$.
The $\alpha=0$ branch contains at most $B(M,d)$ pairs,
each with rejection probability at most $2^{-r}$.
Every order-two Weyl term in the expansion of a qubit
Pauli of weight below $d$ has cost below $d$.
\end{lemma}
The proof is given in Sec.~S3.2 of SM.

\begin{lemma}[Order-four errors]
\label{lem:order-four}
Among local pairs satisfying $4(a_j,b_j)=0$, exactly
$c_h$ have cost one.
At most $W_h(M,d)$ pairs of exact order four have
total cost below $d$; each has rejection probability
at most $2^{-2r}$.
Every order-four Weyl term in the expansion of a qubit
Pauli of weight below $d$ has cost below $d$.
\end{lemma}
The proof is given in Sec.~S3.3 of SM.
The local costs lower-bound the Pauli weight needed to produce
a Weyl term.

\begin{lemma}[Higher-order errors]
\label{lem:higher-order}
For $3\le s\le h$, at most $V_s(M,d)$ pairs of exact
order $2^s$ have total cost below $d$; each has rejection
probability at most $2^{-sr}$.
Every order-$2^s$ Weyl term in the expansion of a qubit
Pauli of weight below $d$ has cost below $d$.
\end{lemma}
The proof is given in Sec.~S3.4 of SM.
Combining these bounds by the union bound gives
$\Pr(\mathsf{reject})\le\Phi_h(M,K,d)$, where
\begin{equation}
 \begin{split}
 \Phi_h(M,K,d)={}&2^{-M}B(2M,d)\\
&+2^{-r}B(M,d)+2^{-2r}W_h(M,d)\\
 &+\sum_{s=3}^h2^{-sr}V_s(M,d).
 \end{split}
 \label{eq:Phi-main}
\end{equation}
The entropy estimate below uses the threshold
$t_\star=2H_2^{-1}(1/2)$, where
\[
 H_2(t)=-t\log_2t-(1-t)\log_2(1-t)
\]
is the binary entropy and the inverse is taken on $[0,1/2]$.
Combining the rejection bounds above with the standard entropy estimate for binary Hamming balls~\cite{Florence_The_1977} yields the following lemma.

\begin{lemma}
\label{lem:entropy-rejection}
Fix $h\ge3$ and $0<t<t_\star$.
There exist constants $c>0$ and $M_0$, depending only on $h,t$,
such that, for every $M\ge M_0$ and $K=d=\lceil tM\rceil$,
\begin{equation}
 \Pr(\mathsf{reject})\le\Phi_h(M,K,d)\le2^{-cM}.
 \label{eq:entropy-rejection-main}
\end{equation}
\end{lemma}
The proof is given in Sec.~S3.5 of SM.
The preceding bounds yield the following computable code family.
\begin{theorem}
\label{thm:exist}
For each fixed $h\ge3$, a computable quantum code family admitting fully addressable
$Z(h)$ gates satisfies
\begin{equation}
 \lim_i\frac{K_i}{N_i}=\frac{t_\star}{h},\qquad
 \liminf_i\frac{D_i}{N_i}\ge\frac{t_\star}{h}.
 \label{eq:exist}
\end{equation}
The defining matrices $\mathsf{B}_i$ and $A_i$ admit deterministic generation in
$2^{O_h(N_i^2)}$ time or randomized generation with exact
verification in expected $2^{O_h(N_i)}$ time, using
polynomial working space.
\end{theorem}
\begin{proof}
Fix $h\ge3$ and write $N=hM$ and $r=M-K$.
Lemma~\ref{lem:selector} gives an isometric, fully addressable
encoding for every admissible matrix pair $(\mathsf B,A)$.
It therefore remains to select pairs with the required distance.

Let $P$ be a nonidentity qubit Pauli of weight below $d$.
The identity coefficient in its Weyl expansion is
$2^{-N}\operatorname{Tr}P=0$.
By Lemmas~\ref{lem:order-two}--\ref{lem:higher-order},
every nonidentity Weyl term appearing in this expansion
has cost below $d$ and is therefore checked by the rejection test.
If no rejection occurs, Lemma~\ref{lem:weyl-compression} gives
$E^\dagger W(a,b)E=0$ for each such term. By linearity,
$E^\dagger PE=0$. Since this holds for every nonidentity Pauli of
weight below $d$, the code has $D\ge d$.
The union bound and the counting lemmas give
$\Pr(\mathsf{reject})\le\Phi_h(M,K,d)$.

For $i\ge3$, choose $t_i$ as the largest integer multiple
of $2^{-i}$ in $(0,1/2)$ satisfying $2H_2(t_i/2)<1$.
This gives a sequence $(t_i)$ converging to $t_\star$,
with $0<t_i<t_\star$.
For each $i$, choose the smallest integer $M_i\ge2^i$ such that
\[
 K_i=d_i=\lceil t_iM_i\rceil,\qquad
 \Phi_h(M_i,K_i,d_i)<\frac12.
\]
Lemma~\ref{lem:entropy-rejection} guarantees that such an
integer exists and can be found by testing successive lengths.
At every selected length, some matrix pair triggers no rejection
and therefore defines a code with $D_i\ge d_i$.
With $N_i=hM_i$,
\[
 \frac{K_i}{N_i}=\frac{d_i}{N_i}
 =\frac{t_i}{h}+O(M_i^{-1}),
\]
which implies Eq.~\eqref{eq:exist}.

These parameter choices are effective.
For rational $t_i$, clearing denominators and exponentiating
reduces the entropy condition to an exact integer comparison.
The finite sums defining $\Phi_h$ can also be evaluated exactly
using rational arithmetic.
Since $2^i\le M_i$, the required numbers have polynomial bit length,
and the parameter search takes polynomial time and space in $N_i$.

For each candidate matrix pair, enumerate the $2^{2hM_i}$ Weyl
pairs and check the rejection conditions for every nonzero pair
of cost below $d_i$. Computing the order and cost of a pair, then
checking these conditions, takes polynomial time and space.
Processing the pairs one at a time therefore gives exact
verification in $2^{O_h(N_i)}$ time and polynomial working space.
There are $2^{O_h(M_i^2)}$ candidate matrix pairs, so exhaustive
search takes $2^{O_h(N_i^2)}$ time with the same space bound.

Uniform admissible matrix pairs can be sampled in polynomial
expected time: a uniformly sampled ring matrix is invertible
with probability greater than $1/4$, as determined by its
reduction modulo $2$.
Since $\Pr(\mathsf{reject})\le\Phi_h(M_i,K_i,d_i)<1/2$,
each uniformly sampled admissible matrix pair triggers no rejection
with probability greater than $1/2$.
Independent repetition therefore requires fewer than two
admissible matrix pairs on average, giving expected time
$2^{O_h(N_i)}$ with exact verification and polynomial working space.
\end{proof}

For $T=Z(3)$, the asymptotic rate is $0.0733519\ldots$, and the
asymptotic relative distance is at least this value. This corresponds
to approximately $13.63$ physical data qubits per logical qubit.
Exact generation selects
the matrices; the resulting encoder has the polynomial
circuit size established in Sec.~\ref{sec:framework}.

The AG construction also applies to $h=1,2$. For $S$,
the counting construction gives $H_2^{-1}(1/2)$.
For $Z$, the order-two count instead gives the root
$\beta_\star$ of
$\beta+H_2(\beta)+\beta\log_2 3=1$, approximately
$0.15$ (Sec.~S3.6 of SM).

\section{Upper bounds from addressability}
\label{sec:bounds}
The addressing matrix $F$ in Eq.~\eqref{eq:address} imposes a classical
distance constraint. Set $B_F=F\bmod2$, and let $s_F$ be the number of
nonzero columns of $B_F$.
\begin{lemma}
\label{lem:parity}
Let an $((N,2^K,D))_2$ quantum code satisfy Eq.~\eqref{eq:address}
with $h\ge1$ and $K\ge1$. Removing the zero columns of $B_F$
gives a generator matrix for a binary linear $[s_F,K,\ge D]$ code.
\end{lemma}
\begin{proof}
Substitute $c=2^{h-1}v$ for $v\in\F_2^K$ in
Eq.~\eqref{eq:address}. Since $Z(h)^{2^{h-1}}=Z$,
\[
 Z^{vB_F}E=EZ^v.
\]
For $v\ne0$, the logical Pauli $Z^v$ is nonscalar.
Therefore $vB_F\ne0$, proving rank $K$, and quantum
distance requires $\wt(vB_F)\ge D$. Removing zero
columns preserves these weights.
\end{proof}
Therefore, the parameters of the quantum code must satisfy the following limits,
\begin{equation}
 2^K\le A_2(s_F,D),\qquad K+D-1\le s_F,
 \label{eq:finiteupper}
\end{equation}
where $A_2(s_F,d)$ is the largest possible number of binary codewords of length $s_F$ and minimum distance at least $d$.
The second inequality is the classical Singleton
bound~\cite{Florence_The_1977}, obtained by puncturing $D-1$
coordinates of the classical code.

\begin{theorem}
\label{thm:upper}
Consider a family of binary quantum
codes with parameters $((N_i,2^{K_i},D_i))_2$ and $N_i\to\infty$.
Suppose for a fixed integer $h\ge1$, each code admits fully addressable transversal powers of
$Z(h)=\diag(1,e^{2\pi i/2^h})$, with addressing matrix $F_i$.
Let $\eta$ be the family's common asymptotic coefficient of rate and relative distance.
Then $ \eta\le c_{\mathrm{up}},$
where $c_{\mathrm{up}}=282652727463/10^{12}$.
If each addressing matrix $F_i$ has the form specified in
Lemma~\ref{lem:selector}, then $ \eta\le c_{\mathrm{up}}/h.$
\end{theorem}
\begin{proof}
{For binary codes of relative distance at least $t\in(0,1/2)$,
the asymptotic rate is at most $\mathcal U(t)$~\cite[Corollary 23]{Omar_Binary_2026}.
The function $\mathcal U$ is continuous, and Sec.~S4 of SM proves
$\mathcal U(c_{\mathrm{up}})<c_{\mathrm{up}}$.}
Pad the parity code of Lemma~\ref{lem:parity} with zero
columns to length $N$. {By continuity, choose $0<t<c_{\mathrm{up}}$
such that $\mathcal U(t)<t$.}
If $\eta > c_{\mathrm{up}}$, these classical codes
eventually have rate and relative distance greater than $t$.
The rate bound then gives
$t < \mathcal U(t)<t$, a contradiction.
Hence $\eta\le c_{\mathrm{up}}$.

{Under the additional condition on the addressing matrices,}
the factor $2^b$ makes every column with $b\ge1$ zero
modulo $2$. Only the column with $b=0$ in each of the
$M$ registers can remain nonzero, so $s_F\le M$.
Pad the resulting classical code with zeros to length
$M=N/h$. Its rate is $K/M=hK/N$, and its relative
distance is at least $D/M=hD/N$.
The same argument gives $h\eta\le c_{\mathrm{up}}$ and hence
$\eta\le c_{\mathrm{up}}/h$.

\end{proof}

{Every code family admitting fully addressable logical $Z(h)$ gates
satisfies $\eta\le c_{\mathrm{up}}$.}
Both constructions in this paper also satisfy Eq.~\eqref{eq:digitform},
so they obey the stronger bound $\eta\le c_{\mathrm{up}}/h$.
Table~\ref{tab:asymptotic} compares the achievable lower
bounds on $\eta$ from both constructions with the universal
upper bound and the stronger bound under
Eq.~\eqref{eq:digitform}, including their values for $T=Z(3)$.

\section{Finite codes and encoding circuits}
\label{sec:finite}
{
We obtain two finite quantum codes with fully addressable transversal
logical $T$ gates: $\mathcal Q_{48}=((48,16,5))_2$ and
$\mathcal Q_{90}=((90,256,7))_2$. They encode $4$ and $8$ logical
qubits into $48$ and $90$ physical qubits, respectively. We find their defining matrices by sampling
candidates from the construction in Sec.~\ref{sec:existence} at $h=3$
and applying its exact rejection tests. For each code, checking every
nonzero Weyl pair of cost below the claimed distance certifies detection
of every Pauli error of lower weight. An explicit nontrivial logical
Pauli of matching weight gives the upper bound, so the distances are
exactly $5$ and $7$. A code of distance $D$ corrects arbitrary errors
on up to $t=\lfloor(D-1)/2\rfloor$ qubits under ideal recovery~\cite{Emanuel_Theory_1997}.
Thus, the two codes
correct arbitrary errors on up to $2$ and $3$ qubits, respectively.}
The rejection {tests} and logical witnesses are given in Sec.~S6.2 of SM;
complete matrices and machine-readable certificates are available from
the authors upon reasonable request.

{By Eq.~\eqref{eq:address} with $h=3$, each logical $T$ gate is
implemented by a transversal layer of physical single-qubit phase gates.
For $\mathcal Q_{48}$, the individual $T$ gates on logical qubits
$1,2,3,4$ have physical gate weights $33$, $29$, $30$, and $36$,
respectively. Here the weight counts the number of nonidentity gates in the transversal layer.}

{An ancilla-free encoding circuit} for $\mathcal Q_{48}$ uses
$466$ two-qubit controlled-phase gates, $24$ two-qubit
SWAP gates, $12$ three-qubit multiplications
$x\mapsto ux\bmod8$ with odd $u$, and $202$ six-qubit
weighted additions $(x,y)\mapsto(x,y+cx\bmod8)$.
Scheduling {these operations} on disjoint qubits in parallel gives a
multi-qubit gate depth of $228$, assuming all-to-all
connectivity and unit duration for each listed {operation}.
Single-qubit gates are excluded from this depth count.
Further circuit details are given in Sec.~S6.3 of SM.

\section{Discussion}
\label{sec:discussion}
Independent phase control can coexist with a nonvanishing
density of protected logical qubits. The cyclic encoding
makes the mechanism explicit: inverse-matrix columns select
logical phases, while classical code distances or quadratic
phase counting provide error detection. This separation
produces quantitative guarantees for general quantum
subspaces and connects their addressability to classical
coding bounds.

The two constructions offer different parameter and
construction-cost tradeoffs. Algebraic-geometric evaluation
matrices give a uniform polynomial-time family. Random
quadratic phases, coupled with exact verification, give a
larger common rate--distance coefficient. The parity-code
converse places both within a common framework and
identifies how the physical digit structure limits the
simultaneous rate and distance.

A natural next step is to combine these selectors with
efficient decoding, low-weight error checks, and scalable
syndrome extraction. For physical implementations, the
simple transversal phase layer should be assessed together
with encoding, recovery, connectivity, and a specified
noise model. The explicit finite encoders provide
concrete starting points for such studies. On the coding
side, tighter error counting, improved inner encodings,
and other physical addressing structures offer routes
to narrowing the remaining gap between construction
guarantees and converse bounds.

\section*{Acknowledgments}
OpenAI's GPT models were used to assist in developing proofs and
searching for optimal code parameters. The authors take full
responsibility for the results and proofs presented in this work.
B.W. and T.L. acknowledge support from the National Natural Science Foundation
of China (Grant No.~12405014).

\section*{Data availability}
The data supporting the findings of this study are available from the
authors upon reasonable request.

\nocite{Serge_Number_1983,Sergey_Magic-state_2012,Adam_Improved_2026}
\bibliographystyle{unsrt}
\bibliography{refs}

\clearpage
\onecolumngrid
\fontsize{12}{14.2}\selectfont
\phantomsection
\pdfbookmark[0]{Supplemental Material}{arxiv-supplement}
\begin{center}
{\fontsize{12}{14.2}\selectfont\bfseries Supplemental Material for:}\par\smallskip
{\fontsize{12}{14.2}\selectfont\bfseries
\parbox{0.7\linewidth}{\centering Substantially Improved Rate and Distance for Addressable Transversal Phase Gates\par}}
\end{center}
\setcounter{section}{0}
\renewcommand{\thesection}{S\arabic{section}}
\setcounter{subsection}{0}
\setcounter{subsubsection}{0}
\setcounter{equation}{0}
\renewcommand{\theequation}{S\arabic{equation}}
\setcounter{theorem}{0}
\renewcommand{\thetheorem}{S\arabic{theorem}}
\setcounter{lemma}{0}
\renewcommand{\thelemma}{S\arabic{lemma}}
\setcounter{proposition}{0}
\renewcommand{\theproposition}{S\arabic{proposition}}
\setcounter{corollary}{0}
\renewcommand{\thecorollary}{S\arabic{corollary}}
\setcounter{definition}{0}
\renewcommand{\thedefinition}{S\arabic{definition}}
\setcounter{algorithm}{0}
\renewcommand{\thealgorithm}{S\arabic{algorithm}}
\setcounter{table}{0}
\renewcommand{\thetable}{S\arabic{table}}
\setcounter{figure}{0}
\renewcommand{\thefigure}{S\arabic{figure}}

\renewcommand{\thesubsection}{\thesection.\arabic{subsection}}
\renewcommand{\thesubsubsection}{\thesubsection.\arabic{subsubsection}}
\makeatletter
\renewcommand{\p@subsection}{}
\renewcommand{\p@subsubsection}{}
\renewcommand{\fnum@table}{Table~\thetable}
\renewcommand{\fnum@figure}{Figure~\thefigure}
\makeatother
\renewcommand{\theHsection}{supp.\arabic{section}}
\renewcommand{\theHsubsection}{\theHsection.\arabic{subsection}}
\renewcommand{\theHsubsubsection}{\theHsubsection.\arabic{subsubsection}}
\providecommand{\theHequation}{}
\renewcommand{\theHequation}{supp.\arabic{equation}}
\providecommand{\theHtheorem}{}
\renewcommand{\theHtheorem}{supp.\arabic{theorem}}
\providecommand{\theHlemma}{}
\renewcommand{\theHlemma}{supp.\arabic{lemma}}
\providecommand{\theHproposition}{}
\renewcommand{\theHproposition}{supp.\arabic{proposition}}
\providecommand{\theHcorollary}{}
\renewcommand{\theHcorollary}{supp.\arabic{corollary}}
\providecommand{\theHdefinition}{}
\renewcommand{\theHdefinition}{supp.\arabic{definition}}
\providecommand{\theHalgorithm}{}
\renewcommand{\theHalgorithm}{supp.\arabic{algorithm}}
\providecommand{\theHtable}{}
\renewcommand{\theHtable}{supp.\arabic{table}}
\providecommand{\theHfigure}{}
\renewcommand{\theHfigure}{supp.\arabic{figure}}

This Supplemental Material provides the proofs and construction details
supporting the main text. Section~\ref{sec:si-encoding} establishes
addressability and derives error-compression identities for the encoding. Sections~\ref{sec:si-ag} and~\ref{sec:si-existence} develop
the polynomial-time AG construction and the counting proof with exact
exponential-time generation, respectively. Section~\ref{sec:si-converse}
derives universal rate--distance upper bounds from addressability and
stronger bounds for cyclic-digit encodings. Section~\ref{sec:si-prior}
compares the parameter guarantees of prior constructions.
Section~\ref{sec:si-examples} verifies the addressable logical gates
and exact distances of the finite examples and describes an ancilla-free
encoder for the $48$-qubit code.

\begingroup
\SMcontentsstyle
\par\medskip
\noindent{\fontsize{14.4}{17.3}\selectfont\bfseries Contents\par}
\medskip
\makeatletter
\l@section{\hyperref[arxiv:si-contents-1]{\numberline{\ref*{arxiv:si-contents-1}}Encoding and logical phase gates}}{\hyperref[arxiv:si-contents-1]{\pageref*{arxiv:si-contents-1}}}
\l@section{\hyperref[arxiv:si-contents-2]{\numberline{\ref*{arxiv:si-contents-2}}Polynomial-time algebraic-geometric construction}}{\hyperref[arxiv:si-contents-2]{\pageref*{arxiv:si-contents-2}}}
\l@subsection{\hyperref[arxiv:si-contents-3]{\numberline{\ref*{arxiv:si-contents-3}}Ring lifting and the outer code}}{\hyperref[arxiv:si-contents-3]{\pageref*{arxiv:si-contents-3}}}
\l@subsection{\hyperref[arxiv:si-contents-4]{\numberline{\ref*{arxiv:si-contents-4}}Inner encoding and concatenation}}{\hyperref[arxiv:si-contents-4]{\pageref*{arxiv:si-contents-4}}}
\l@subsection{\hyperref[arxiv:si-contents-5]{\numberline{\ref*{arxiv:si-contents-5}}Tower parameters and rate--distance bounds}}{\hyperref[arxiv:si-contents-5]{\pageref*{arxiv:si-contents-5}}}
\l@subsection{\hyperref[arxiv:si-contents-6]{\numberline{\ref*{arxiv:si-contents-6}}Polynomial-time generation}}{\hyperref[arxiv:si-contents-6]{\pageref*{arxiv:si-contents-6}}}
\l@section{\hyperref[arxiv:si-contents-7]{\numberline{\ref*{arxiv:si-contents-7}}Counting proof and certified exponential construction}}{\hyperref[arxiv:si-contents-7]{\pageref*{arxiv:si-contents-7}}}
\l@subsection{\hyperref[arxiv:si-contents-8]{\numberline{\ref*{arxiv:si-contents-8}}Compression and a common probability bound}}{\hyperref[arxiv:si-contents-8]{\pageref*{arxiv:si-contents-8}}}
\l@subsection{\hyperref[arxiv:si-contents-9]{\numberline{\ref*{arxiv:si-contents-9}}Order-two errors}}{\hyperref[arxiv:si-contents-9]{\pageref*{arxiv:si-contents-9}}}
\l@subsection{\hyperref[arxiv:si-contents-10]{\numberline{\ref*{arxiv:si-contents-10}}Order-four errors}}{\hyperref[arxiv:si-contents-10]{\pageref*{arxiv:si-contents-10}}}
\l@subsection{\hyperref[arxiv:si-contents-11]{\numberline{\ref*{arxiv:si-contents-11}}Higher-order errors}}{\hyperref[arxiv:si-contents-11]{\pageref*{arxiv:si-contents-11}}}
\l@subsection{\hyperref[arxiv:si-contents-12]{\numberline{\ref*{arxiv:si-contents-12}}The total rejection bound}}{\hyperref[arxiv:si-contents-12]{\pageref*{arxiv:si-contents-12}}}
\l@subsection{\hyperref[arxiv:si-contents-13]{\numberline{\ref*{arxiv:si-contents-13}}Exponential-time construction algorithm}}{\hyperref[arxiv:si-contents-13]{\pageref*{arxiv:si-contents-13}}}
\l@section{\hyperref[arxiv:si-contents-14]{\numberline{\ref*{arxiv:si-contents-14}}Rate-distance upper bounds from addressability}}{\hyperref[arxiv:si-contents-14]{\pageref*{arxiv:si-contents-14}}}
\l@section{\hyperref[arxiv:si-contents-15]{\numberline{\ref*{arxiv:si-contents-15}}Quantitative comparison with prior constructions}}{\hyperref[arxiv:si-contents-15]{\pageref*{arxiv:si-contents-15}}}
\l@section{\hyperref[arxiv:si-contents-16]{\numberline{\ref*{arxiv:si-contents-16}}Finite examples and their exact certificates}}{\hyperref[arxiv:si-contents-16]{\pageref*{arxiv:si-contents-16}}}
\l@subsection{\hyperref[arxiv:si-contents-17]{\numberline{\ref*{arxiv:si-contents-17}}Addressable transversal \texorpdfstring{$T$}{T} gates}}{\hyperref[arxiv:si-contents-17]{\pageref*{arxiv:si-contents-17}}}
\l@subsection{\hyperref[arxiv:si-contents-18]{\numberline{\ref*{arxiv:si-contents-18}}Verification of code distances}}{\hyperref[arxiv:si-contents-18]{\pageref*{arxiv:si-contents-18}}}
\l@subsection{\hyperref[arxiv:si-contents-19]{\numberline{\ref*{arxiv:si-contents-19}}Ancilla-free encoding circuit}}{\hyperref[arxiv:si-contents-19]{\pageref*{arxiv:si-contents-19}}}
\makeatother
\endgroup
\clearpage
\section{Encoding and logical phase gates}
\label{arxiv:si-contents-1}
\label{sec:si-encoding}

A binary quantum code $((N,2^K,D))_2$ encodes $K$ logical qubits
into $N$ physical qubits and has distance $D$. Its rate and
relative distance are $\mathsf R=K/N$ and $\mathsf d=D/N$.

A code projector $P$ detects an operator $O$ when $POP$ is a
scalar multiple of $P$. Equivalently, for an encoding isometry
$E$ with $P=EE^\dagger$, the compression $E^\dagger OE$ must be a
scalar multiple of the logical identity~\cite{Emanuel_Theory_1997}. The distance is the minimum
Pauli weight for which this fails. For $N_i\to\infty$, write
\[
 \mathsf R_\infty=\liminf_i K_i/N_i,\qquad
 \mathsf d_\infty=\liminf_i D_i/N_i,\qquad
 \eta=\min\{\mathsf R_\infty,\mathsf d_\infty\}.
\]
Lower asymptotic targets $\mathsf R_0,\mathsf d_0$ mean
$\mathsf R_\infty\ge \mathsf R_0$ and
$\mathsf d_\infty\ge \mathsf d_0$.

For an integer $h\ge1$, the physical phase gate is
\[
 Z(h)=\diag(1,e^{2\pi i/2^h}).
\]
Set
\[
 \nu=2^h,\qquad \rh=\Z/\nu\Z.
\]
The phase level $h$ is fixed in every asymptotic family.
Throughout, $\log$ denotes the base-two logarithm and $\ln$ the natural logarithm. Vectors are row vectors.
We use addressability on a fixed encoding, as in
Ref.~\cite[Definition 1.1]{Tongyin_Asymptotically_2026}.

\begin{definition}[Full addressability]
\label{def:si-address}
An encoding isometry $E:(\mathbb C^2)^{\otimes K}\to
(\mathbb C^2)^{\otimes N}$ is fully addressable at level $h$ if there
is $F\in\rh^{K\times N}$ such that
\begin{equation}
 \left(\bigotimes_{j=1}^N Z(h)^{(cF)_j}\right)E
 =E\left(\bigotimes_{\ell=1}^K Z(h)^{c_\ell}\right)
 \quad\text{for every }c\in\rh^K.
 \label{eq:si-address}
\end{equation}
Physical powers, including inverse powers, are allowed. The encoding
is fixed for every logical exponent vector $c$.
\end{definition}

For $y\in\rh$, its digit representation is
$\dig(y)=(y_0,\ldots,y_{h-1})$, where
$y= {\sum_{t=0}^{h-1}2^t y_t}$ and $ {y_t}\in\{0,1\}$; this convention is
applied coordinatewise. Each ring coordinate is represented by
a register of $h$ physical qubits.
{We also use $\dig$ for the linear isometry
$\ket y\mapsto\ket{\dig(y)}=\ket{y_0,\ldots,y_{h-1}}$,
applied independently to every ring register.}
{We identify each ring register with its $h$-qubit digit
representation and use the same notation for states and operators.}
On one such register define
\[
 X_\nu(a)\ket y=\ket{y+a},\qquad
 Z_\nu(b)\ket y=e^{2\pi iby/\nu}\ket y.
\]
For $M$ ring registers and $a,b\in\rh^M$, define
$W(a,b)=\bigotimes_{j=1}^M X_\nu(a_j)Z_\nu(b_j)$. These operators form an
orthogonal operator basis, including for composite $\nu$:
$\Tr(W(a,b)^\dagger W(a',b'))=\nu^M\mathbf1_{(a,b)=(a',b')}$.
This follows by first taking the trace of the shift, which vanishes
unless $a=a'$, and then summing the remaining cyclic character.
Every operator $O$ therefore has the Weyl expansion
\referenceedit{\begin{equation}
 O=\nu^{-M}\sum_{a,b}\Tr\!\left(W(a,b)^\dagger O\right)W(a,b).
 \label{eq:si-weyl-expansion}
\end{equation}}
The number of registers on which $W(a,b)$ acts nontrivially is
$\wt_{\mathrm s}(a,b)=|\supp(a)\cup\supp(b)|$.

The following encoding will be used in both constructions. Choose
$\mathsf B\in\GL_M(\rh)$, integers $r,K$ with $r+K\le M$, and a symmetric
matrix $A$ over $\rh$. Let $G$ and $L$ comprise the first $r$ and
next $K$ rows of $\mathsf B$.
Put $\tau=e^{\pi i/\nu}$ and define
\begin{equation}
 E_{\mathsf B,A}\ket{x}
 =\nu^{-r/2}\sum_{v\in\rh^r}
       \tau^{\,yAy^{\mathsf T}}\ket{\dig(y)},\qquad
 y=vG+xL,\quad x\in\{0,1\}^K,
 \label{eq:si-encoding}
\end{equation}
where in the quadratic phase, we use integer representatives in
$\{0,\ldots,\nu-1\}$ for all entries of $A$ and $y$.
The output length is $N=hM$.

\begin{lemma}%
\label{lem:si-encoding}
Eq.~\eqref{eq:si-encoding} defines an isometry, and its addressing
matrix is
\begin{equation}
 F_{\ell,(j, {t})}
 =2^{ {t}}(\mathsf B^{-1})_{j,r+\ell}\pmod{\nu},
 \quad 1\le\ell\le K,\quad 1\le j\le M,\quad 0\le t<h.
 \label{eq:si-F}
\end{equation}
{Let $C_1$ be the row span of the first $r+K$ rows of
$\mathsf B$ over $\rh$. For $a,b\in\rh^M$ and
$x,x'\in\{0,1\}^K$, the matrix element} ${\langle x'|E_{\mathsf B,A}^\dagger
 W(a,b)E_{\mathsf B,A}|x\rangle}$
is nonzero if and only if
\begin{equation}
 a\in C_1,\qquad
 G(b-aA)^{\mathsf T}=0,\qquad
 x'-x=(a\mathsf B^{-1})_{r+1:r+K}
 \quad\text{over }\rh.
 \label{eq:si-compression}
\end{equation}
\end{lemma}
\begin{proof}
The first $r+K$ rows of an invertible ring matrix are a free family.
Consequently $v\mapsto vG+xL$ is injective, and two retained binary
labels give disjoint cosets. Each coset has $\nu^r$ elements, and
each amplitude has modulus $\nu^{-r/2}$, proving orthonormality. If
$Q_A\ket y=\tau^{yAy^{\mathsf T}}\ket y$, direct calculation gives
\referenceedit{\begin{equation}
 Q_A^\dagger W(a,b)Q_A
 =\tau^{-aAa^{\mathsf T}}W(a,b-aA).
 \label{eq:si-phase-conjugation}
\end{equation}}
Write $\gamma=b-aA$ and $q=a\mathsf B^{-1}$. Equality of supports requires
$q_{r+K+1:M}=0$ and $q_{r+1:r+K}=x'-x$. Subject to these conditions,
the remaining matrix element is
\begin{equation}
 \begin{split}
 \langle x'|E_{\mathsf B,A}^{\dagger}W(a,b)E_{\mathsf B,A}|x\rangle
 &=\tau^{-aAa^{\mathsf T}}e^{2\pi i\gamma(xL)^{\mathsf T}/\nu}\\
 &\quad\times
   \nu^{-r}\sum_{v\in\rh^r}e^{2\pi i\gamma(vG)^{\mathsf T}/\nu}.
 \end{split}
 \label{eq:si-compression-value}
\end{equation}
The normalized character sum equals one if $G\gamma^{\mathsf T}=0$
and zero otherwise: a nonzero coefficient in any coordinate produces
a vanishing geometric sum over that coordinate. This proves
Eq.~\eqref{eq:si-compression} over the cyclic ring.

Let $z_\ell$ be row $r+\ell$ of $\mathsf B^{-\mathsf T}$. Then
$Gz_\ell^{\mathsf T}=0$ and $Lz_\ell^{\mathsf T}=e_\ell^{\mathsf T}$.
Consequently $\bigotimes_jZ_\nu(z_{\ell,j})$ has phase
$e^{2\pi ix_\ell/\nu}$ on every term of the encoded state. Finally,
\referenceedit{\begin{equation}
 Z_\nu(a)=\bigotimes_{ {t}=0}^{h-1}Z(h)^{2^{t}a}
 \label{eq:si-digit-phase}
\end{equation}}
under the digit representation. This proves~Eq.~\eqref{eq:si-F} and
Eq.~\eqref{eq:si-address}; the quadratic phase commutes with these gates.
\end{proof}

The \emph{cyclic-digit class} consists of the encodings in
Eq.~\eqref{eq:si-encoding} with the addressing form in Eq.~\eqref{eq:si-F}. The upper
bound for this class will use only that form, and therefore also
applies to other encodings admitting such a matrix.
The matrices compactly specify the encoding and its gates.
For fixed $h$, reversible modular elimination
implements $y\mapsto y\mathsf B$ with $O(M^2)$ fixed-size arithmetic gates~\referenceedit{\cite[Secs.~III--IV]{Erik_Stabilizer_2005}}.
The diagonal $Q_A$ uses $O(M^2)$ gates, each acting on at most two
$h$-qubit registers.
Hence an encoder of polynomial size is available once the matrices
have been selected; the cost of selecting them is treated separately.

\section{Polynomial-time algebraic-geometric construction}
\label{arxiv:si-contents-2}
\label{sec:si-ag}

We give the ring-lifting and outer-code arguments used in
Sec.~III~A of the main text, then extend its inner encoding
from eight registers to any positive even number $m$.
This extension allows the comparison of fixed square binary
fields. We finish with explicit tower parameters and the
deterministic construction algorithm.

\subsection{Ring lifting and the outer code}
\label{arxiv:si-contents-3}

For a fixed partition of coordinates into blocks, block weight
counts the blocks containing a nonzero coordinate.
Write $d_{\mathrm{bl}}(C)$ for the minimum nonzero block weight
of a nonzero linear code $C$.

\begin{lemma}%
\label{lem:si-free}
Let $\mathsf B_2\in\GL_M(\F_2)$ and lift its zero--one entries to
$\mathsf B\in\GL_M(\rh)$. For $1\le s<M$, let $B_s$ and $C_s$ be
the binary and ring row spans of the first $s$ rows. Then
$ d_{\mathrm{bl}}(C_s)=d_{\mathrm{bl}}(B_s),
 d_{\mathrm{bl}}(C_s^\perp)=d_{\mathrm{bl}}(B_s^\perp),$
where the ring dual is $ C_s^\perp=\{z\in\rh^M:cz^{\mathsf T}=0
 \text{ for every }c\in C_s\},$
with the dot product computed modulo $2^h$.
\end{lemma}
\begin{proof}
Since $\mathsf B\bmod 2$ is invertible, $\det\mathsf B$ is
odd and hence a unit in $\rh$. Thus $\mathsf B$ is invertible
over $\rh$.

Let $G$ consist of the first $s$ rows of $\mathsf B$, and
write $G_2=G\bmod 2$. These matrices generate $C_s$ and
$B_s$, respectively.

Take any nonzero codeword $c\in C_s$ and write $c=uG$.
Factor the largest common power of two from the coefficients:
$u=2^ku'$, where $0\le k<h$ and at least one entry of $u'$
is odd. Set
\[
 w=u'G,\qquad b=w\bmod 2=(u'\bmod 2)G_2.
\]
The rows of $G_2$ are linearly independent, so $b$ is a
nonzero codeword of $B_s$.
Whenever $b_j=1$, the coordinate $w_j$ is odd, and therefore
$c_j=2^kw_j$ is nonzero modulo $2^h$.
Thus every nonzero block of $b$ is also nonzero in $c$, giving
\[
 \mathrm{wt}_{\mathrm{bl}}(c)
 \ge \mathrm{wt}_{\mathrm{bl}}(b)
 \ge d_{\mathrm{bl}}(B_s).
\]
Since this holds for every nonzero $c\in C_s$,
$d_{\mathrm{bl}}(C_s)\ge d_{\mathrm{bl}}(B_s)$.

For the reverse inequality, choose a nonzero codeword
$b=\alpha G_2$ of minimum block weight in $B_s$.
Regard the entries of $\alpha$ as $0$ or $1$ in $\rh$ and set
\[
 c_*=2^{h-1}\alpha G\in C_s.
\]
Because $\alpha G\bmod 2=b$, multiplication by $2^{h-1}$
gives
\[
 (c_*)_j=
 \begin{cases}
  0, & b_j=0,\\
  2^{h-1}, & b_j=1.
 \end{cases}
\]
Hence $c_*$ is nonzero and has exactly the same nonzero
blocks as $b$. Consequently,
\[
 d_{\mathrm{bl}}(C_s)
 \le \mathrm{wt}_{\mathrm{bl}}(c_*)
 =d_{\mathrm{bl}}(B_s),
\]
which proves the first equality.

For the dual codes, let $H$ consist of the last $M-s$ rows
of $\mathsf B^{-\mathsf T}$.
The identity
$\mathsf B(\mathsf B^{-\mathsf T})^{\mathsf T}=I$
shows that these rows form a basis of $C_s^\perp$.
Moreover,
\[
 \mathsf B^{-\mathsf T}\bmod 2
 =(\mathsf B\bmod 2)^{-\mathsf T},
\]
so the rows of $H\bmod 2$ form a basis of $B_s^\perp$.
The preceding argument uses only the linear independence
of the reduced rows, and therefore also applies to $H$.
It gives
$d_{\mathrm{bl}}(C_s^\perp)=d_{\mathrm{bl}}(B_s^\perp)$.
\end{proof}

Let $X/\F_{2^m}$ have genus $g$, a rational point $P_\infty$,
and $n$ other rational points with sum $\mathcal D$. For
$2g-2<a_0<a_1<n$, set
\[
 \mathcal A_b=C_{\mathrm{L}}(\mathcal D,a_bP_\infty),\qquad b=0,1.
\]
Expand them in a fixed binary basis to obtain $B_0\subset B_1$,
keeping the $m$ coordinates of each field symbol in one block.
Lift a complete basis adapted to $B_0\subset B_1$ to
$\mathsf B_{\mathrm{out}}\in\GL_{mn}(\rh)$. Its first
$r_{\mathrm{out}}=m(a_0+1-g)$ rows form $G$, and its next
$K=m(a_1-a_0)$ rows form $L$.
These dimensions follow from Riemann--Roch
\cite{Henning_Algebraic_2009}.

\begin{lemma}
\label{lem:si-outer}
For $G$ and $L$ defined above, the encoding
\begin{equation}
 E_{\mathrm{out}}\ket{x}
 =\nu^{-r_{\mathrm{out}}/2}
  \sum_{v\in\rh^{r_{\mathrm{out}}}}\ket{vG+xL},
 \qquad x\in\{0,1\}^K,
 \label{eq:si-outer-encoding}
\end{equation}
detects every operator acting on fewer than
\begin{equation}
 d_*=\min\{n-a_1,a_0+2-2g\}
 \label{eq:si-outer-block-distance}
\end{equation}
blocks, where each block consists of the $m$ ring registers
associated with one field coordinate.
\end{lemma}

\begin{proof}
The rows of $G$ and $L$ belong to a basis of $\rh^{mn}$,
so distinct coefficient pairs $(v,x)$ give distinct
computational-basis states. The normalization in
Eq.~\eqref{eq:si-outer-encoding} therefore makes
$E_{\mathrm{out}}$ an isometry.
If $d_*=1$, the assertion concerns only scalar multiples
of the identity and is immediate. Assume henceforth that
$d_*\ge2$.

We first obtain the two classical distance bounds needed
for the proof. Let $\theta_1,\ldots,\theta_m$ be the basis
used to expand the field symbols, and let
$\theta_1^*,\ldots,\theta_m^*$ be its trace-dual basis\referenceedit{~\cite[Chap.~2]{Rudolf_Introduction_1994}}:
$ \Tr(\theta_j\theta_\ell^*)=\delta_{j\ell},$
where $\Tr$ is the field trace to $\F_2$.
For $\xi=\sum_j\xi_j\theta_j$ and
$\zeta=\sum_j\zeta_j\theta_j^*$, this gives
$ \Tr(\xi\zeta)=\sum_j\xi_j\zeta_j$.
Applying this identity coordinatewise shows that expanding
$\mathcal A_0^\perp$ in the trace-dual basis gives binary
vectors orthogonal to $B_0$.
The resulting binary code and $B_0^\perp$ both have dimension
$mn-r_{\mathrm{out}}$, so they are equal.
In either basis, a field symbol is zero exactly when its
binary block is zero. The standard AG distance bounds
\referenceedit{\cite[Chap.~2]{Henning_Algebraic_2009}} thus imply
\[
 d_{\mathrm{bl}}(B_1)\ge n-a_1,\qquad
 d_{\mathrm{bl}}(B_0^\perp)\ge a_0+2-2g.
\]

By Lemma~\ref{lem:si-free}, every nonzero ring linear
combination of the rows of $G$ and $L$ has at least
$n-a_1$ nonzero blocks. Likewise, every nonzero ring vector
$b$ satisfying $Gb^{\mathsf T}=0$ has at least
$a_0+2-2g$ nonzero blocks.

Now consider a nonidentity Weyl operator $W(a,b)$ acting
on fewer than $d_*$ blocks. Both $a$ and $b$ have block
weight less than $d_*$.

If $a\ne0$, a nonzero matrix element between encoded basis
states would require
\[
 vG+xL+a=v'G+x'L
\]
for some $v,v',x,x'$. Equivalently,
\[
 a=(v'-v)G+(x'-x)L.
\]
This would make $a$ a nonzero ring linear combination of
the rows of $G$ and $L$ with fewer than $n-a_1$ nonzero
blocks, contradicting the distance bound. Hence
\[
 E_{\mathrm{out}}^\dagger W(a,b)E_{\mathrm{out}}=0.
\]

If $a=0$, then $b\ne0$. The second distance bound implies
$Gb^{\mathsf T}\ne0$.
The operator $W(0,b)$ is diagonal, and distinct encoded
basis states have disjoint computational-basis supports,
so all off-diagonal matrix elements vanish.
\referenceedit{Eq.~\eqref{eq:si-compression-value} with $A=0$ and
$a=0$ gives the diagonal matrix elements}
\[
 \langle x|E_{\mathrm{out}}^\dagger
 W(0,b)E_{\mathrm{out}}|x\rangle
 =
 \frac{e^{2\pi i(xL)b^{\mathsf T}/\nu}}
      {\nu^{r_{\mathrm{out}}}}
 \sum_{v\in\rh^{r_{\mathrm{out}}}}
 e^{2\pi i vGb^{\mathsf T}/\nu}.
\]
Since $Gb^{\mathsf T}$ has a nonzero coordinate, summing
over the corresponding coordinate of $v$ gives a
vanishing geometric sum. Thus
$E_{\mathrm{out}}^\dagger W(0,b)E_{\mathrm{out}}=0$ as well.

\referenceedit{By Eq.~\eqref{eq:si-weyl-expansion}, every operator
$O$ on fewer than $d_*$ blocks is a linear combination of Weyl
operators on those same blocks.} All nonidentity terms vanish between encoded
states, while the identity term gives
$E_{\mathrm{out}}^\dagger E_{\mathrm{out}}=I$.
Consequently,
\[
 E_{\mathrm{out}}^\dagger O E_{\mathrm{out}}
 =\lambda_O I,
\]
which is precisely the error-detection condition.
\end{proof}

\subsection{Inner encoding and concatenation}
\label{arxiv:si-contents-4}

{The outer bound \referenceedit{of Lemma~\ref{lem:si-outer}} counts affected
field-coordinate blocks. We now encode each block to detect
single-register errors while preserving the phase operations
used for addressability.}

For positive even $m$, set $\epsilon_a=1$ for $a\le m/2$ and
$\epsilon_a=-1$ otherwise. Thus $\sum_a\epsilon_a=0$.
For $u\in\rh^m$, define
\begin{equation}
 J_m\ket u=\nu^{-1/2}\sum_{t\in\rh}
 \ket{u_1+t,\ldots,u_m+t,\ t,\
             t-\sum_{a=1}^m\epsilon_a u_a}.
 \label{eq:si-inner-encoding}
\end{equation}
All arithmetic is modulo $\nu$. This maps $m$ ring registers
to $m+2$ registers and reduces to the inner encoding of
Sec.~III~A of the main text at $m=8$.

\begin{lemma}%
\label{lem:si-inner}
The map $J_m$ is an isometry with distance two, measured in
ring registers. For $z\in\rh^m$, put
$\widetilde z=(z_1,\ldots,z_m,-\sum_a z_a,0)$. Then
\begin{equation}
 \left(\bigotimes_{j=1}^{m+2}Z_\nu(\widetilde z_j)\right)J_m
 =J_m\left(\bigotimes_{a=1}^{m}Z_\nu(z_a)\right).
 \label{eq:si-inner-selector}
\end{equation}
\end{lemma}
\begin{proof}
For fixed $u$ and $t$, write the computational-basis label in Eq. \eqref{eq:si-inner-encoding} as
\[
 w=w(u,t)=(w_1,\ldots,w_{m+2})
 =
 \left(u_1+t,\ldots,u_m+t,\,
 t,\,
 t-\sum_{a=1}^m\epsilon_a u_a\right).
\]
Its $(m+1)$st coordinate determines $t$, and then
$u_a=w_a-t$ recovers each input coordinate.
Thus distinct pairs $(u,t)$ give distinct basis states.
Each encoded state contains $\nu$ such states with amplitude
$\nu^{-1/2}$, so the encoded basis states are orthonormal
and $J_m$ is an isometry.

We next prove detection of every single-register error.
Since $\sum_a\epsilon_a=0$, every basis label appearing
in an encoded state satisfies
\[
 \sum_{a=1}^m\epsilon_a w_a-w_{m+1}+w_{m+2}=0
 \pmod{\nu}.
\]
Consider a nonidentity Weyl operator $Q$ acting on one
output register, with shift parameter $a$ and phase
parameter $b$.

If $a\ne0$, the shift changes exactly one coordinate of $w$.
Every coefficient in the displayed constraint is $1$ or $-1$,
so its left-hand side changes by $a$ or $-a$, which is
nonzero modulo $\nu$. The shifted basis labels therefore
cannot occur in any encoded state. Consequently,
\[
 J_m^\dagger QJ_m=0.
\]

If $a=0$, then $b\ne0$ and $Q$ is a phase operator on some
register $j$. Distinct encoded basis states have disjoint
supports, so all off-diagonal matrix elements vanish.
For the diagonal elements, every output coordinate satisfies
$w_j(u,t)=w_j(u,0)+t$, giving
\[
 \langle u|J_m^\dagger QJ_m|u\rangle
 =
 \frac{e^{2\pi i b w_j(u,0)/\nu}}{\nu}
 \sum_{t=0}^{\nu-1}e^{2\pi i bt/\nu}
 =0.
\]
The last equality follows because $b\ne0$ modulo $\nu$.
Hence $J_m^\dagger QJ_m=0$ for every nonidentity
single-register Weyl operator.
Since these Weyl operators together with the identity
span all single-register operators, every such operator
$O$ satisfies
\[
 J_m^\dagger OJ_m=\lambda_O I.
\]
Thus the register distance is at least two.

To prove Eq.~\eqref{eq:si-inner-selector}, evaluate the
physical phase exponent on each basis label:
\[
 \begin{aligned}
 \widetilde z\,w(u,t)^{\mathsf T}
 &=\sum_{a=1}^m z_a(u_a+t)
   -\left(\sum_{a=1}^m z_a\right)t\\
 &=\sum_{a=1}^m z_a u_a.
 \end{aligned}
\]
This exponent is independent of $t$.
Every term in $J_m\ket u$ therefore acquires the same
phase, equal to the action of
$\bigotimes_{a=1}^m Z_\nu(z_a)$ on $\ket u$.
This proves the claimed identity.

Finally, choose $z=(1,0,\ldots,0)$.
The corresponding physical operator applies $Z_\nu(1)$
to output register $1$ and $Z_\nu(-1)$ to output register
$m+1$, acting trivially elsewhere.
By Eq.~\eqref{eq:si-inner-selector}, its action on the
encoded inputs is $Z_\nu(1)$ on the first input register.
This two-register operator has a nonscalar logical action
and therefore fails the error-detection condition.
A code of register distance at least three would detect
every two-register operator, so the distance is at most two.
{Together with single-register error detection,
this proves that the register distance is exactly two.}
\end{proof}

{Apply $J_m$ to every outer block and use
$\dig$ to express its output registers as qubits. For a Pauli error,
\referenceedit{Lemma~\ref{lem:si-inner} implies that}
any inner block with nonscalar compression contains at least two
affected qubits. Lemma~\ref{lem:si-outer} then gives the
physical-qubit distance bound, while
Eq.~\eqref{eq:si-inner-selector} preserves the logical phase operations.}

\begin{lemma}%
\label{lem:si-concatenation}
For even $m$, {the binary quantum code defined by}
$E=\dig\circ J_m^{\otimes n}\circ E_{\mathrm{out}}$ has full
addressability in cyclic-digit form and parameters
\begin{equation}
 N=h(m+2)n,\qquad K=m(a_1-a_0),\qquad
 D\ge2\min\{n-a_1,a_0+2-2g\}.
 \label{eq:si-AG-finite}
\end{equation}
\end{lemma}
\begin{proof}
Each stage of the encoding is an isometry.
The inner encoding replaces each block of $m$ ring registers
by $m+2$ registers, each represented by $h$ physical qubits.
Thus $N=h(m+2)n$, while the number of logical qubits remains
$K=m(a_1-a_0)$.

Let $d_*$ be defined in
Eq.~\eqref{eq:si-outer-block-distance}, and consider a physical
Pauli operator $Q$ of weight $w<2d_*$.
Group its action by inner blocks.
Within any block containing at most one affected physical
qubit, the error acts on at most one ring register.
Lemma~\ref{lem:si-inner} therefore gives a scalar multiple
of the identity under $J_m^\dagger(\cdot)J_m$.
A nonscalar action can remain only in blocks containing
at least two affected physical qubits, of which there are
at most $\lfloor w/2\rfloor<d_*$.
Lemma~\ref{lem:si-outer} detects every operator on these
remaining blocks, so
\[
 E^\dagger QE=\lambda_Q I.
\]
Thus every physical Pauli operator of weight below $2d_*$
is detected, proving $D\ge2d_*$.

For addressability, let $z_\ell^{\mathsf T}$ be column
$r_{\mathrm{out}}+\ell$ of $\mathsf B_{\mathrm{out}}^{-1}$.
\referenceedit{The selector construction in
Lemma~\ref{lem:si-encoding} gives}
\[
 (vG+xL)z_\ell^{\mathsf T}=x_\ell,
\]
so the register phases with exponents $z_\ell$ implement
$Z(h)$ on logical qubit $\ell$.

Partition $z_\ell$ into blocks $z_{\ell,p}\in\rh^m$ and set
\[
 \widetilde z_{\ell,p}
 =\left(z_{\ell,p},-\sum_{j=1}^m z_{\ell,p,j},0\right).
\]
Eq.~\eqref{eq:si-inner-selector} preserves this logical
action through the inner encoding.
\referenceedit{The digit-phase identity in Eq.~\eqref{eq:si-digit-phase}
then gives the physical addressing matrix}
\begin{equation}
 \begin{gathered}
 F_{\ell,(p,j,b)}
 =2^b\widetilde z_{\ell,p,j}\pmod\nu,
 \end{gathered}
 \label{eq:si-concatenated-selector}
\end{equation}
where $ 1\le\ell\le K, 1\le p\le n,
 1\le j\le m+2, 0\le b<h.$
For any $c\in\rh^K$, combining these phase operations
with exponents $c_\ell$ gives physical exponents $cF$
and logical exponents $c$, proving
Eq.~\eqref{eq:si-address}.

Finally, introduce an auxiliary input $v\in\rh$ and consider
the ring-linear map
\begin{equation}
 (u_1,\ldots,u_m,t,v)\longmapsto
 \left(u_1+t,\ldots,u_m+t,\,
 t,\,
 t-\sum_{a=1}^m\epsilon_a u_a+v\right).
 \label{eq:si-inner-linear-map}
\end{equation}
Output coordinate $m+1$ determines $t$, after which $u$
and $v$ are uniquely recovered. Thus
Eq.~\eqref{eq:si-inner-linear-map} defines a unitary
permutation of the computational basis. Preparing the
two auxiliary registers in a uniform superposition and
$\ket0$, respectively, gives
\[
 \begin{aligned}
 &\ket u\otimes
 \left(\nu^{-1/2}\sum_{t\in\rh}\ket t\right)
 \otimes\ket0\\
 &\quad\longmapsto
 \nu^{-1/2}\sum_{t\in\rh}
 \left|u_1+t,\ldots,u_m+t,\,
 t,\,
 t-\sum_{a=1}^m\epsilon_a u_a\right\rangle\\
 &\quad=J_m\ket u,
 \end{aligned}
\]
where the last equality is the definition in
Eq.~\eqref{eq:si-inner-encoding}.
By linearity, the same implementation applies to any
input state.
Composing these maps with $\mathsf B_{\mathrm{out}}$ and
ordering the inputs as uniform, binary logical, and zero
inputs gives Eq.~\eqref{eq:si-encoding} with
\[
 A=0,\qquad M=(m+2)n,\qquad r=r_{\mathrm{out}}+n,
\]
and $K$ logical qubits.
\end{proof}

\subsection{Tower parameters and rate--distance bounds}
\label{arxiv:si-contents-5}
\label{sec:si-ag-parameters}

Fix $h\ge3$. We use the finite parameters in
Eq.~\eqref{eq:si-AG-finite} to choose the AG degrees and derive
the asymptotic guarantees. For a fixed positive even $m$,
consider curves over $\F_{2^m}$ with $n_i\to\infty$ evaluation
points and $g_i/n_i\to\kappa$. Choose constants
$2\kappa<\beta_0<\beta_1<1$ and set
$a_{b,i}=\lfloor\beta_b n_i\rfloor$ for $b=0,1$.
These degrees satisfy $2g_i-2<a_{0,i}<a_{1,i}<n_i$ for all
sufficiently large $i$. Since rounding changes each normalized
degree by less than $1/n_i$, Eq.~\eqref{eq:si-AG-finite} gives
\begin{equation}
 \begin{aligned}
 \lim_i\frac{K_i}{N_i}
 &=\frac{m(\beta_1-\beta_0)}{h(m+2)},\\
 \liminf_i\frac{D_i}{N_i}
 &\ge\frac{2\min\{1-\beta_1,\beta_0-2\kappa\}}{h(m+2)}.
 \end{aligned}
 \label{eq:si-AG-tradeoff}
\end{equation}

For the family constructed in Sec.~III~A of the main text, take $m=8$ and the
Garcia--Stichtenoth tower over $\F_{256}$
\cite{Arnaldo_On_1996,Maria_The_2006}:
\[
 F_1=\F_{256}(x_1),\qquad F_{i+1}=F_i(x_{i+1}),\qquad
 x_{i+1}^{16}+x_{i+1}=\frac{x_i^{16}}{x_i^{15}+1}.
\]
At level $i\ge1$, the rational places above
$x_1=\alpha\in\F_{256}\setminus\F_{16}$ provide
$n_i=240\cdot16^{i-1}$ evaluation points. The unique pole
$P_{\infty,i}$ of $x_1$ is rational and lies outside this set.
The genus is
\cite{Arnaldo_On_1996,Maria_The_2006}
\[
 g_i=\begin{cases}
 (16^{i/2}-1)^2,&i\text{ even},\\
 (16^{(i+1)/2}-1)(16^{(i-1)/2}-1),&i\text{ odd}.
 \end{cases}
\]
Hence $g_i\le16^i=n_i/15$ and $g_i/n_i\to1/15$.

At $m=8$ and $\kappa=1/15$, balancing the rate and the two
distance terms in Eq.~\eqref{eq:si-AG-tradeoff} requires
\[
 4(\beta_1-\beta_0)=1-\beta_1=\beta_0-2/15.
\]
Solving gives $\beta_0=70/135$ and $\beta_1=83/135$, so choose
\begin{equation}
 a_{0,i}=\left\lfloor\frac{70n_i}{135}\right\rfloor,\qquad
 a_{1,i}=\left\lfloor\frac{83n_i}{135}\right\rfloor.
 \label{eq:si-poly-degrees}
\end{equation}
The genus bound and rounding give
\[
 \begin{aligned}
 a_{1,i}-a_{0,i}&\ge\frac{13n_i}{135}-1,\\
 n_i-a_{1,i}&\ge\frac{52n_i}{135},\\
 a_{0,i}+2-2g_i&\ge\frac{52n_i}{135}+1.
 \end{aligned}
\]
Since $n_i\ge240$, these inequalities also imply
$2g_i-2<a_{0,i}<a_{1,i}<n_i$ for every $i\ge1$.
Substituting into Eq.~\eqref{eq:si-AG-finite}, with
$N_i=10hn_i$, yields the finite-length guarantees
\[
 \frac{K_i}{N_i}\ge\frac{52}{675h}-\frac4{5hn_i},\qquad
 \frac{D_i}{N_i}\ge\frac{52}{675h}.
\]
Moreover, $(a_{1,i}-a_{0,i})/n_i\to13/135$, so
\[
 \lim_i\frac{K_i}{N_i}=\frac{52}{675h},\qquad
 \liminf_i\frac{D_i}{N_i}\ge\frac{52}{675h},
\]
as stated in Sec.~III~A of the main text. The encoding and addressing
matrices are generated by the polynomial-time procedure in
Sec.~\ref{sec:si-ag-algorithm}.

To obtain the rate--distance tradeoff, keep the same tower
and choose
\[
 \beta_0=\frac2{15}+\tau_{\mathrm d},\qquad
 \beta_1=1-\tau_{\mathrm d},\qquad
 0<\tau_{\mathrm d}<\frac{13}{30}.
\]
The two distance terms are then equal, and
Eq.~\eqref{eq:si-AG-tradeoff} becomes
\[
 \mathsf R_\infty=\frac4{5h}
       \left(\frac{13}{15}-2\tau_{\mathrm d}\right),\qquad
 \mathsf d_\infty\ge\frac{\tau_{\mathrm d}}{5h}.
\]
Thus a positive target pair $(\mathsf R_0,\mathsf d_0)$ is
attainable whenever
\begin{equation}
 \frac54h\mathsf R_0+10h\mathsf d_0<\frac{13}{15}.
 \label{eq:si-AG-region}
\end{equation}
Indeed, this condition makes the interval
\[
 5h\mathsf d_0<\tau_{\mathrm d}
 <\frac{13}{30}-\frac{5h}{8}\mathsf R_0
\]
nonempty. Choosing a rational $\tau_{\mathrm d}$ in this
interval and rounding the degrees as above makes both
asymptotic guarantees strictly exceed their targets.
The balanced choice is $\tau_{\mathrm d}=52/135$; at $h=3$
its common guarantee is $52/2025$.

Finally, we compare the common guarantees supplied by
Eq.~\eqref{eq:si-AG-tradeoff} for fixed square binary fields.
For a fixed genus ratio $\kappa$, both bounds can reach a
coefficient $\eta>0$ only if
\[
 \beta_1-\beta_0\ge\frac{h(m+2)}m\eta,\qquad
 1-\beta_1\ge\frac{h(m+2)}2\eta,\qquad
 \beta_0-2\kappa\ge\frac{h(m+2)}2\eta.
\]
The left sides sum to $1-2\kappa$, so the largest common
guarantee from these bounds is
\begin{equation}
 \eta_{\mathrm{AG},m}
 =\frac{m(1-2\kappa)}{h(m+1)(m+2)},\qquad 2\kappa<1.
 \label{eq:si-inner-optimized}
\end{equation}
It is attained by taking $\beta_0=2\kappa+ {\tau_{\mathrm d}}$ and
$\beta_1=1- {\tau_{\mathrm d}}$, where
$ {\tau_{\mathrm d}}=m(1-2\kappa)/(2(m+1))$; all three inequalities then
hold with equality.

The Drinfeld--Vl\u{a}du\c{t} bound
\cite{Serge_Number_1983} gives
$\kappa\ge(2^{m/2}-1)^{-1}$.
For $m=4,6,8$, the resulting upper bounds on
$h\eta_{\mathrm{AG},m}$ are $2/45$, $15/196$, and $52/675$,
respectively; $m=2$ cannot satisfy $2\kappa<1$.
For every even $m\ge10$,
\[
 h\eta_{\mathrm{AG},m}
 <\frac{m}{(m+1)(m+2)}
 \le\frac5{66}<\frac{52}{675}.
\]
The $\F_{256}$ tower attains $\kappa=1/15$ and therefore
gives the largest common guarantee obtained from
Eq.~\eqref{eq:si-AG-tradeoff} with the inner encodings $J_m$
over fixed square binary fields.

\subsection{Polynomial-time generation}
\label{arxiv:si-contents-6}
\label{sec:si-ag-algorithm}

For the balanced family, the input is a fixed phase level $h$
and a tower level $i$. The output is an encoding matrix, its
addressing matrix $F$, and the certified parameters in
Eq.~\eqref{eq:si-AG-finite} at $m=8$.
\begin{itemize}
\item Construct the $\F_{256}$ tower curve, its $n_i$ evaluation
places and $P_{\infty,i}$, and choose the degrees in
Eq.~\eqref{eq:si-poly-degrees}.
\item Compute the two AG evaluation matrices~\cite{Kenneth_A_2001}.
Expand each field coordinate into eight binary coordinates,
and complete a basis adapted to the nested binary codes.
\item Lift the basis to $\mathsf B_{\mathrm{out}}$ over $\rh$
\referenceedit{as in Lemma~\ref{lem:si-free}}.
Its first $r_{\mathrm{out}}=8(a_{0,i}+1-g_i)$ rows and next
$K_i=8(a_{1,i}-a_{0,i})$ rows define $G$ and $L$.
Compute the required columns of $\mathsf B_{\mathrm{out}}^{-1}$.
\item For each block, use the invertible ring-linear
transformation defined in Eq.~\eqref{eq:si-inner-linear-map} with $m=8$.
Extend the outer transformation defined by $\mathsf B_{\mathrm{out}}$
by the identity on the auxiliary registers, and then apply these
block transformations.
Compute $F$ using Eq.~\eqref{eq:si-concatenated-selector}.
\end{itemize}
The resulting cyclic encoding has $A=0$, $M_i=10n_i$,
$r_i=r_{\mathrm{out}}+n_i$, and $N_i=10hn_i$.
The deterministic algorithm of Shum et al.~\cite{Kenneth_A_2001}
computes the evaluation matrices in
$O(n_i^3(\log n_i)^3)$ field operations for this fixed field.
Binary expansion, basis completion, ring-matrix inversion and
composition are polynomial in $n_i$ for fixed $h$.
These operations therefore generate the encoding and addressing
matrices deterministically in polynomial time in $N_i$.
{For each inner block, initialize two auxiliary registers
in $\ket0$. Apply Hadamards to the $h$ qubits of the first register
to prepare the uniform $t$ input. Then implement
Eq.~\eqref{eq:si-inner-linear-map} with $m=8$ by modular additions.}
Modular elimination and the inner linear maps also give an
$O_h(N_i^2)$-size encoding circuit.

\section{Counting proof and certified exponential construction}
\label{arxiv:si-contents-7}
\label{sec:si-existence}

We prove existence by showing that a random encoding has positive
probability of satisfying a sufficient condition for distance at
least $d$. We derive this condition from the action of Weyl errors,
and then use counting and entropy estimates to obtain the asymptotic
rate and distance guarantees. The same condition provides exact
verification for the exponential-time construction.

Throughout this section, $1\le K<M$, $1\le d\le M$, $r=M-K$,
and $\nu=2^h$. Choose $\mathsf B$ uniformly from $\GL_M(\rh)$
and $A$ independently and uniformly among symmetric matrices.
\referenceedit{Use these matrices in the encoding of
Eq.~\eqref{eq:si-encoding}.}
The first $r$ and last $K$ rows of $\mathsf B$ are $G$ and $L$.
A nonzero Weyl pair $(a,b)$ has exact order $2^s$ if
$2^s(a,b)=0$ but $2^{s-1}(a,b)\ne0$. Equivalently, all coordinates
are divisible by $2^{h-s}$, and at least one resulting quotient
is odd. We use the counting functions
\begin{align}
 B(n,d)&=\sum_{w=1}^{d-1}\binom nw,\notag\\
 V_s(M,d)&=\sum_{w=1}^{d-1}(4^s-1)^w\binom Mw,\notag\\
 W_h(M,d)&=\sum_{w=1}^{d-1}[z^w]
       (1+c_hz+(15-c_h)z^2)^M,\qquad
 c_h=\begin{cases}10,&h=2,3,\\6,&h\ge4.\end{cases}
 \label{eq:si-volumes}
\end{align}
The last function is used for $h\ge2$; empty sums are zero.

\subsection{Compression and a common probability bound}
\label{arxiv:si-contents-8}
\label{sec:si-compression-proof}

Here we derive the rejection test and a probability bound
for each fixed Weyl pair. We start by applying
Lemma~\ref{lem:si-encoding} with $r+K=M$ to determine when a Weyl
operator has a nonzero matrix element between encoded states.

\begin{lemma}
\label{lem:si-compression}
For $x,x'\in\{0,1\}^K$, the matrix element
$\langle x'|E^\dagger W(a,b)E|x\rangle$ is nonzero if and only if
\[
 G(b-aA)^{\mathsf T}=0,\qquad
 (a\mathsf B^{-1})_{r+1:M}=x'-x.
\]
\end{lemma}
\begin{proof}
Apply Lemma~\ref{lem:si-encoding} with $r+K=M$.
Since $\mathsf B$ is invertible, its rows span $\rh^M$,
so $C_1=\rh^M$ and the condition $a\in C_1$ is automatic.
The remaining two conditions are exactly those stated here.
\end{proof}

For $h\ge2$, reject a candidate matrix pair if some nonzero Weyl
pair of cost below $d$ satisfies $G(b-aA)^{\mathsf T}=0$ and, at
exact order two, also satisfies $(a\mathsf B^{-1})_{r+1:M}=0$.
{The costs add over registers and are summarized in Table~\ref{tab:si-rejection-summary}, together with the counting and probability bounds proved below.}
Let
$\mathsf{reject}$ denote the event that at least one such pair exists.
At exact order two, the logical shift has entries in
$\{0,2^{h-1}\}$ and can equal a difference of binary labels only
when it vanishes.
{For pairs of exact order at least four, we omit the logical-shift condition to simplify the sufficient test for $D\ge d$.}
Thus Lemma~\ref{lem:si-compression} implies that
no rejection gives $E^\dagger W(a,b)E=0$ for every tested pair.
Algorithm~\ref{alg:si-search} implements this test.

At exact order two, the cost equals the physical Pauli weight of $W(a,b)$.
At higher orders, the cost is a lower bound on the weight of any qubit Pauli whose Weyl expansion contains the pair with a nonzero coefficient.

\begin{table}[htbp]
\centering
\small
\renewcommand{\arraystretch}{1.2}
\caption{Costs and bounds for nonzero Weyl pairs of cost below $d$, for $h\ge2$.
The last column bounds the probability that a fixed pair triggers rejection.
The bounds are proved in Lemmas~\ref{lem:si-order-two}--\ref{lem:si-higher-order}.}
\label{tab:si-rejection-summary}
\begin{tabular}{@{}llcc@{}}
\toprule
Pair class & Cost & Count bound & Probability bound\\
\midrule
Order $2$, $\alpha\ne0$ & $\wt(\alpha)+\wt(\beta)$ & $B(2M,d)$ & $2^{-M}$\\
Order $2$, $\alpha=0$ & $\wt(\beta)$ & $B(M,d)$ & $2^{-r}$\\
Order $4$ & Sum of local costs $0,1,2$ & $W_h(M,d)$ & $2^{-2r}$\\
Order $2^s$, $3\le s\le h$ & Number of nonzero registers & $V_s(M,d)$ & $2^{-sr}$\\
\bottomrule
\end{tabular}
\smallskip

\parbox{\textwidth}{For order two, $(a,b)=2^{h-1}(\alpha,\beta)$ with $\alpha,\beta\in\F_2^M$.
For order four, write each local pair as $(a_j,b_j)=2^{h-2}(\alpha_j,\beta_j)$ with $\alpha_j,\beta_j\in\Z/4\Z$.
Its cost is zero for $(0,0)$, one for labels in the set defined by Eq.~\eqref{eq:si-localsets}, and two otherwise.}
\end{table}

The following
estimate supplies the probability bound for each exact order.

\begin{lemma}
\label{lem:si-order-probability}
For a fixed Weyl pair $(a,b)$ of exact order $2^s$, $1\le s\le h$,
\[
 \Pr\!\left[G(b-aA)^{\mathsf T}=0\right]\le2^{-sr}.
\]
\end{lemma}
\begin{proof}
Divide $(a,b)$ by $2^{h-s}$ and reduce all matrices modulo $2^s$.
The reductions of $\mathsf B$ and $A$ remain independent and
uniform among invertible and symmetric matrices, respectively.
Write the reduced pair as $(a',b')$;
at least one of its coordinates is odd.

Suppose $a'_j$ is odd, hence invertible modulo $2^s$.
For any target vector $w$, choose the symmetric submatrix
$(A_{k\ell})_{k,\ell\ne j}$ arbitrarily. For each $\ell\ne j$,
the equation
\[
 a'_j A_{j\ell}
 =w_\ell-\sum_{k\ne j}a'_k A_{k\ell}
\]
uniquely determines $A_{j\ell}=A_{\ell j}$.
The remaining equation $(a'A)_j=w_j$ then uniquely determines
$A_{jj}$. Thus exactly $2^{sM(M-1)/2}$ symmetric matrices
satisfy $a'A=w$ for each $w$, so $a'A$ is uniformly distributed.
Since $A$ is independent of $\mathsf B$, the vector
$\gamma=b'-a'A$ remains uniform for every fixed $\mathsf B$.

Invertibility of $\mathsf B$ implies that
$\mathsf B\gamma^{\mathsf T}$ is also uniform.
The condition $G\gamma^{\mathsf T}=0$ fixes its first $r$
coordinates to zero and leaves its last $K$ coordinates
arbitrary. Therefore
\[
 \Pr\!\left[G\gamma^{\mathsf T}=0\mid\mathsf B\right]
 =\frac{2^{sK}}{2^{sM}}
 =2^{-sr}.
\]

Otherwise, all coordinates of $a'$ are even, while $b'$ has
an odd coordinate. For each fixed $A$, the vector
$\gamma=b'-a'A$ therefore has an odd coordinate.

For uniformly chosen invertible $\mathsf B$, the vector
$\mathsf B\gamma^{\mathsf T}$ is uniform among all vectors
with at least one odd coordinate. Indeed, invertible row
operations can transform any such vector into the first
coordinate vector. Thus any two such outputs are related
by an invertible matrix, and left multiplication by this
matrix gives a bijection between the corresponding choices
of $\mathsf B$.

There are $2^{sM}-2^{(s-1)M}$ possible outputs, excluding
those whose coordinates are all even.
The condition $G\gamma^{\mathsf T}=0$ requires the first $r$
coordinates to vanish, leaving $K$ coordinates with at least
one odd entry. Hence
\[
 \Pr\!\left[G\gamma^{\mathsf T}=0\mid A\right]
 =\frac{2^{sK}-2^{(s-1)K}}
        {2^{sM}-2^{(s-1)M}}
 =2^{-sr}\frac{1-2^{-K}}{1-2^{-M}}
 <2^{-sr},
\]
where the last inequality uses $K<M$.

This proves the bound in both cases.
\end{proof}

\subsection{Order-two errors}
\label{arxiv:si-contents-9}
\label{sec:si-order-two}

We now count Weyl errors of exact order two with weight below $d$
and bound their rejection probabilities.
The following lemma bounds the number of order-two Weyl errors
of cost below $d$ and their rejection probabilities, separately
for $\alpha\ne0$ and $\alpha=0$, where
$a=2^{h-1}\alpha$ with $\alpha\in\{0,1\}^M$.

\begin{lemma}
\label{lem:si-order-two}
For $h\ge2$, write each Weyl pair of exact order two as
$(a,b)=2^{h-1}(\alpha,\beta)$ with
$\alpha,\beta\in\{0,1\}^M$.
Among pairs with error weight
$\wt(\alpha)+\wt(\beta)<d$, the cases $\alpha\ne0$
and $\alpha=0$ contain at most $B(2M,d)$ and $B(M,d)$
pairs, with rejection probabilities per pair at most
$2^{-M}$ and $2^{-r}$, respectively.
\end{lemma}

\begin{proof}
The operators $X_\nu(2^{h-1})$ and $Z_\nu(2^{h-1})$ act as
$X$ on the highest digit and $Z$ on the lowest, respectively.
These digits are distinct for $h\ge2$, so $W(a,b)$ is itself
a qubit Pauli of weight $\wt(\alpha)+\wt(\beta)$.
Pauli orthogonality implies that an order-two Weyl term
appearing in a qubit Pauli $P$ has cost $\wt(P)$.
Thus every such term is tested when $\wt(P)<d$.
Counting binary vectors of weight below $d$ in $2M$ coordinates
gives $B(2M,d)$; when $\alpha=0$, only $M$ coordinates remain,
giving $B(M,d)$.

For $\alpha\ne0$, write
$\overline{\mathsf B}=\mathsf B\bmod2$ and
$q=\alpha\overline{\mathsf B}^{-1}$.
At exact order two, Lemma~\ref{lem:si-compression}
implies that nonzero compression requires
$(a\mathsf B^{-1})_{r+1:M}=0$.
Since $a=2^{h-1}\alpha$, this condition is equivalent to
$q_{r+1:M}=0$.
Uniformity of $\overline{\mathsf B}$ among invertible binary
matrices makes $q$ uniform on $\mathbb F_2^M\setminus\{0\}$.
Exactly $2^r-1$ of these vectors have their last $K$ coordinates
zero. Hence
\[
 \Pr\!\left[(a\mathsf B^{-1})_{r+1:M}=0\right]
 =\frac{2^r-1}{2^M-1}<2^{-K}.
\]
For every fixed $\mathsf B$, the proof of
Lemma~\ref{lem:si-order-probability}, with $s=1$ and
$\alpha\ne0$, gives
\[
 \Pr\!\left[G(b-aA)^{\mathsf T}=0\mid\mathsf B\right]
 =2^{-r}.
\]
Since the shift condition depends only on $\mathsf B$,
\[
 \begin{aligned}
 &\Pr\!\left[
 G(b-aA)^{\mathsf T}=0,\,
 (a\mathsf B^{-1})_{r+1:M}=0
 \right]\\
 &\qquad
 =2^{-r}\frac{2^r-1}{2^M-1}
 <2^{-M}.
 \end{aligned}
\]
For $\alpha=0$, we have $a=0$, so rejection requires only
$Gb^{\mathsf T}=0$. Lemma~\ref{lem:si-order-probability}
bounds this probability by $2^{-r}$.

\end{proof}

\subsection{Order-four errors}
\label{arxiv:si-contents-10}
\label{sec:si-order-four}

We next bound the number of order-four Weyl terms that can appear
in qubit Pauli errors of weight below $d$. The following lemma
derives this bound from single-qubit expansions and bounds each
term's rejection probability using
Lemma~\ref{lem:si-order-probability}.

\begin{lemma}
\label{lem:si-order-four}
For $h\ge2$, there is a set of at most $W_h(M,d)$ Weyl pairs
of exact order four containing every such pair with a nonzero
coefficient in the Weyl expansion of a qubit Pauli of weight
below $d$. Each pair in this set triggers rejection with
probability at most $2^{-2r}$.
\end{lemma}
\begin{proof}
Write a local Weyl pair as
$(a,b)=2^{h-2}(\alpha,\beta)$, with
$\alpha,\beta\in\mathbb Z_4$. Let $\mathcal A_h$ be the set
of nonzero labels $(\alpha,\beta)$ that occur with nonzero
coefficient in a single-qubit Pauli expansion. For a Pauli
$P_j$ acting on digit $j$, where $0\le j<h$,
\referenceedit{Eq.~\eqref{eq:si-weyl-expansion} with $M=1$ gives the coefficient}
\[
 \frac1\nu\Tr\!\left(W(a,b)^\dagger P_j\right)
 =\frac1\nu\sum_{y=0}^{\nu-1}
 e^{-2\pi i\beta y/4}\langle y+a|P_j|y\rangle.
\]
We evaluate it using
$y=\sum_{\ell=0}^{h-1}2^\ell y_\ell$, with $y_\ell\in\{0,1\}$.

For $Z_j$, a nonzero coefficient requires $\alpha=0$.
Since $Z_j\ket y=(-1)^{y_j}\ket y$, the coefficient then equals
\[
 \frac1\nu\left(1-e^{-2\pi i\beta2^j/4}\right)
 \prod_{\substack{0\le\ell<h\\\ell\ne j}}
 \left(1+e^{-2\pi i\beta2^\ell/4}\right).
\]
It is nonzero exactly for $(j,\beta)=(0,2),(1,1),(1,3)$:
for $j\ge2$ the first factor vanishes, and the cases $j=0,1$
follow by substituting $\beta\in\mathbb Z_4$.
Since every $Z_j$ has $\alpha=0$, these cases give the
Weyl label $(\alpha,\beta)=(0,2)$ for $Z_0$, and the
labels $(\alpha,\beta)=(0,1),(0,3)$ for $Z_1$.

For $X_j$ and $Y_j$ with $j<h-1$, the possible shifts are
$a=\pm2^j$. The condition $a\equiv0\pmod{2^{h-2}}$ therefore
forces $j=h-2$, giving $\alpha=1$ or $3$.
For either shift, $y_j$ is fixed. Summing over the remaining
bits gives, up to a nonzero phase and the factor $\nu^{-1}$,
\[
 \prod_{\substack{0\le\ell<h\\\ell\ne h-2}}
 \left(1+e^{-2\pi i\beta2^\ell/4}\right).
\]
The $X_j$ and $Y_j$ coefficients differ only by a nonzero
phase, so both have the same allowed labels.
For $h=2$, this product is $1+(-1)^\beta$; for $h=3$, it is
$2(1+e^{-\pi i\beta/2})$; for $h\ge4$, it is
$2^{h-3}(1+e^{-\pi i\beta/2})(1+(-1)^\beta)$.
Consequently, the allowed values are
\[
 \alpha\in\{1,3\},\qquad
 \beta\in
 \begin{cases}
  \{0,2\},&h=2,\\
  \{0,1,3\},&h=3,\\
  \{0\},&h\ge4.
 \end{cases}
\]

At the highest digit, $X_{h-1}=X_\nu(2^{h-1})$ gives
$(\alpha,\beta)=(2,0)$. Since
$Y_{h-1}=iX_{h-1}Z_{h-1}$, its allowed phase labels are those
of $Z_{h-1}$ found above. They exist only when $h=2$ and
$\beta\in\left\{1,3\right\}$, giving the additional labels $(2,1),(2,3)$.

Combining the three cases gives
\begin{equation}
\mathcal A_h=
\begin{cases}
 \{(0,1),(0,2),(0,3),(2,0),(2,1),(2,3)\}
 \,\cup\bigl(\{1,3\}\times\{0,2\}\bigr),&h=2,\\[2pt]
 \{(0,1),(0,2),(0,3),(2,0)\}
 \,\cup\bigl(\{1,3\}\times\{0,1,3\}\bigr),&h=3,\\[2pt]
 \bigl(\{0\}\times\{1,2,3\}\bigr)
 \,\cup\bigl(\{1,2,3\}\times\{0\}\bigr),&h\ge4.
\end{cases}
\label{eq:si-localsets}
\end{equation}
The groups with $\alpha=0$, $\alpha=2$, and
$\alpha\in\{1,3\}$ are disjoint. Counting their labels yields
\[
 c_h=|\mathcal A_h|=
 \begin{cases}
  3+3+4=10,&h=2,\\
  3+1+6=10,&h=3,\\
  3+1+2=6,&h\ge4.
 \end{cases}
\]

For counting, assign local weight zero to the identity,
one to pairs in $\mathcal A_h$, and two to the remaining
nonzero pairs. A pair outside $\mathcal A_h$ cannot occur
in a one-qubit Pauli expansion. Hence the assigned weight
of any local term is at most the physical Pauli weight
producing it. Since Weyl expansions factor over registers,
every term in a Pauli error of weight below $d$ has total
assigned weight below $d$.

Each register has one pair of weight zero, $c_h$ pairs of
weight one, and $15-c_h$ pairs of weight two. Consequently,
the number of pairs of exact order four with total assigned
weight below $d$ is at most
\[
 \sum_{w=1}^{d-1}[z^w]
 \left(1+c_hz+(15-c_h)z^2\right)^M
 =W_h(M,d).
\]
This count may also include pairs of order two, which only
increases the upper bound. For every pair of exact order four,
Lemma~\ref{lem:si-order-probability} with $s=2$ gives rejection
probability at most $2^{-2r}$.

\end{proof}

\subsection{Higher-order errors}
\label{arxiv:si-contents-11}
\label{sec:si-higher-order}

For $3\le s\le h$, the following lemma bounds the number of
Weyl errors of exact order $2^s$ supported on fewer than $d$
registers and their rejection probabilities.

\begin{lemma}
\label{lem:si-higher-order}
For $3\le s\le h$, give a Weyl pair cost equal to its number of
nonzero registers. At most $V_s(M,d)$ pairs of exact order $2^s$
have cost below $d$, each with rejection probability at most
$2^{-sr}$. Every order-$2^s$ Weyl term in a qubit Pauli of weight
below $d$ has cost below $d$.
\end{lemma}
\begin{proof}
A qubit Pauli of weight below $d$ touches fewer than $d$ registers.
Its Weyl expansion is the identity on every untouched register,
so each term has cost below $d$. For support size $w$, choose
the registers in $\binom Mw$ ways and a nonzero local pair in
$(4^s-1)^w$ ways. Summing over $1\le w<d$ gives $V_s(M,d)$.
This count includes pairs of smaller order and therefore bounds the number of pairs of exact order $2^s$ with cost below $d$.
For those pairs,
Lemma~\ref{lem:si-order-probability} gives the rejection bound
$2^{-sr}$.
\end{proof}

\subsection{The total rejection bound}
\label{arxiv:si-contents-12}
\label{sec:si-entropy-proof}

Combining the three error classes gives the finite rejection bound.

\begin{lemma}
\label{lem:si-finite}
For $h\ge2$, the rejection probability is at most
\begin{equation}
 \begin{split}
 \Phi_h(M,K,d):={}&2^{-M}B(2M,d)+2^{-r}B(M,d)+2^{-2r}W_h(M,d)+\sum_{s=3}^h2^{-sr}V_s(M,d).
 \end{split}
 \label{eq:si-Phi}
\end{equation}
For $h=1$, the corresponding bound is
\begin{equation}
 \Phi_1(M,K,d):=2^{-r}\sum_{w=1}^{d-1}3^w\binom Mw.
 \label{eq:si-Phi1}
\end{equation}
In either case, if $\Phi_h<1$, some matrix pair triggers no rejection and defines
a fully addressable code with $N=hM$ and $D\ge d$.
\end{lemma}
\begin{proof}
For $h\ge2$, the union bound applied to
Lemmas~\ref{lem:si-order-two}--\ref{lem:si-higher-order}
gives $\Pr(\mathsf{reject})\le\Phi_h$.
If $\Phi_h<1$, the probability of no rejection is positive.
Choose a matrix pair for which no rejection occurs.

Let $P$ be a nonidentity qubit Pauli of weight below $d$.
Its identity Weyl coefficient is $\nu^{-M}\Tr P=0$
\referenceedit{by Eq.~\eqref{eq:si-weyl-expansion}}.
Lemmas~\ref{lem:si-order-two}--\ref{lem:si-higher-order}
ensure that every remaining Weyl term with nonzero coefficient
satisfies the weight restriction for its order.
Since no rejection occurs, Lemma~\ref{lem:si-compression}
gives $E^\dagger W(a,b)E=0$ for each such term.
Linearity then yields $E^\dagger PE=0$, proving $D\ge d$.
Lemma~\ref{lem:si-encoding} gives full addressability
with $N=hM$.

For $h=1$, each register is one qubit, and the Pauli weight of
$(a,b)$ is $|\supp(a)\cup\supp(b)|$.
Reject a candidate if some nonzero Pauli pair of weight below $d$
satisfies
$G(b-aA)^{\mathsf T}=0$.
There are $3^w\binom Mw$ Pauli pairs of weight $w$, so
Lemma~\ref{lem:si-order-probability} with $s=1$ and the
union bound give $\Pr(\mathsf{reject})\le\Phi_1$.
If $\Phi_1<1$, some matrix pair triggers no rejection.
For this pair, every nonzero Pauli pair of weight below $d$
satisfies $G(b-aA)^{\mathsf T}\ne0$.
Lemma~\ref{lem:si-compression} therefore gives
$E^\dagger W(a,b)E=0$, proving $D\ge d$.
Lemma~\ref{lem:si-encoding} gives full addressability with $N=M$.
\end{proof}

We next show that the rejection bound decays exponentially
when $K$ and $d$ grow proportionally to $M$.
Let
\[
 H_2(t)=-t\log_2t-(1-t)\log_2(1-t),
 \qquad t_\star=2H_2^{-1}(1/2),
\]
where the inverse is taken on $[0,1/2]$.

\begin{lemma}
\label{lem:si-entropy}
Fix $h\ge2$ and $0<t<t_\star$. There exist constants $c>0$ and
$M_0$, depending only on $h,t$, such that, for every $M\ge M_0$
and $K=d=\lceil tM\rceil$,
\[
 \Phi_h(M,K,d)\le 2^{-cM}.
\]
\end{lemma}
\begin{proof}
Put $K=d=\lceil tM\rceil$ and $r=M-K$. Then
$d-1<tM$ and $r>(1-t)M-1$.
We bound each of the four contributions in Eq.~\eqref{eq:si-Phi}
separately and show that each decays exponentially in $M$.

The standard binary Hamming-ball estimate
\cite{Florence_The_1977} gives
\[
 B(2M,d)\le2^{2MH_2(t/2)},\qquad
 B(M,d)\le2^{MH_2(t)}.
\]

For the first term, corresponding to order two with $\alpha\ne0$,
\[
 2^{-M}B(2M,d)
 \le2^{-M[1-2H_2(t/2)]}.
\]
The decay rate $1-2H_2(t/2)$ is positive because $t<t_\star$.

For the second term, corresponding to order two with $\alpha=0$,
the same Hamming-ball estimate, together with
$r>(1-t)M-1$, gives
\[
 2^{-r}B(M,d)
 \le 2\cdot 2^{-M[1-t-H_2(t)]}.
\]
To verify positivity of this decay rate, put
$\bar t=p/q=221/1000$. The exact integer comparisons
\[
 (2q)^{2q}>2^q p^p(2q-p)^{2q-p},\qquad
 q^q<2^{q-p}p^p(q-p)^{q-p}
\]
give $2H_2(\bar t/2)>1$ and $\bar t+H_2(\bar t)<1$.
Thus $t<t_\star<\bar t$, and monotonicity of $t+H_2(t)$
on $(0,1/2)$ gives $1-t-H_2(t)>0$.

For the third term, corresponding to order four, let
$P_h(z)=1+c_hz+(15-c_h)z^2$
\referenceedit{be the polynomial defining $W_h$ in Eq.~\eqref{eq:si-volumes}}.
For $0<x<1$ and $w<d$, we have $x^w\ge x^{tM}$.
Since the coefficients of $P_h(z)^M$ are nonnegative,
\[
 W_h(M,d)
 =\sum_{w=1}^{d-1}[z^w]P_h(z)^M
 \le x^{-tM}P_h(x)^M.
\]
Choose $x=1/36$. For both possible values $c_h=6,10$,
\[
 P_h(1/36)\le1+\frac{10}{36}+\frac5{36^2}
 =\frac{1661}{1296}.
\]
Consequently,
\[
 2^{-2r}W_h(M,d)
 \le4\,2^{-M[2-2t-\log_2(1661/1296)-t\log_2 36]}.
\]
The exact comparison
\[
 1661^q36^p<2^{2(q-p)}1296^q
\]
shows that $2\bar t+\log_2(1661/1296)+\bar t\log_2 36<2$.
Monotonicity in $t$ then gives a positive decay rate
for every $t<t_\star$.

For the fourth contribution, fix $3\le s\le h$.
Since $w\le d-1<tM$ in this sum,
\[
 V_s(M,d)
 \le(4^s-1)^{tM}B(M,d)
 \le2^{M[H_2(t)+t\log_2(4^s-1)]}.
\]
It follows that
\[
 2^{-sr}V_s(M,d)
 \le2^s\,2^{-M[s-st-H_2(t)-t\log_2(4^s-1)]}.
\]
Using $H_2(t)\le1$, $\log_2(4^s-1)<2s$, and
$t<\bar t<2/9$, we obtain
\[
 st+H_2(t)+t\log_2(4^s-1)
 <3st+1<\frac{2s}{3}+1\le s.
\]
Thus every higher-order term also has a positive decay rate.
This contribution is absent when $h=2$.

Finally, let $\delta>0$ be the minimum of the decay rates
in these bounds. It depends only on $h,t$.
The prefactors sum to
$1+2+4+\sum_{s=3}^h2^s=2^{h+1}-1$, so
\[
 \Phi_h(M,K,d)\le(2^{h+1}-1)2^{-\delta M}.
\]
Set $c=\delta/2$ and choose $M_0$ large enough that $K<M$
and $2^{h+1}-1\le2^{cM}$ for every $M\ge M_0$.
Then $\Phi_h(M,K,d)\le2^{-cM}$, as claimed.
\end{proof}

\subsection{Exponential-time construction algorithm}
\label{arxiv:si-contents-13}
\label{sec:si-exact-parameters}

For fixed $h\ge2$ and each $i\ge3$, let $t_i$ be the largest
integer multiple of $2^{-i}$ satisfying
$0<t_i<1/2$ and $2H_2(t_i/2)<1$.
These values satisfy $0<t_i<t_\star$ and form a sequence
converging to $t_\star$.
To determine $t_i$ exactly, write $t_i=\mu_i/q_i$, where
$q_i=2^i$ and $\mu_i\in\{1,\ldots,q_i/2-1\}$, and test
$2H_2(t_i/2)<1$ using the equivalent integer comparison
\[
 (2q_i)^{2q_i}<2^{q_i}\mu_i^{\mu_i}(2q_i-\mu_i)^{2q_i-\mu_i}.
\]
With this $t_i$ fixed, search $M=2^i,2^i+1,\ldots$ and set
$K=d=\lceil t_iM\rceil$ at each length; the first triple
satisfying $\Phi_h(M,K,d)<1/2$ is denoted by $(M_i,K_i,d_i)$.
For each fixed $i$, Lemma~\ref{lem:si-entropy} guarantees that
the search stops at a finite length $M_i$.
The additional condition $M_i\ge2^i$ ensures that the selected
code lengths $N_i=hM_i$ tend to infinity as $i\to\infty$.

By Lemma~\ref{lem:si-finite}, these parameters admit codes with $D_i\ge d_i$.
Since $K_i=d_i=\lceil t_iM_i\rceil$ and $t_i\to t_\star$,
\begin{equation}
    \frac{K_i}{N_i}=\frac{d_i}{N_i}
    =\frac{t_i}{h}+O(M_i^{-1})
    \longrightarrow\frac{t_\star}{h}.
    \label{eq:si-counting-rate-limit}
\end{equation}
Thus,
\begin{equation}
    \mathsf R_\infty=\frac{t_\star}{h},\qquad
    \mathsf d_\infty\ge\frac{t_\star}{h}.
    \label{eq:si-counting-endpoint}
\end{equation}

Since $H_2(t_i/2)$ increases with $t_i$, the search over $\mu_i$
can use binary search. Its operands have $O(i2^i)$ bits,
polynomial in $M_i$.
Evaluation of $\Phi_h$ uses integer polynomial arithmetic with
$O_h(M_i)$-bit coefficients. Searching the dyadic grid and scanning
lengths through $M_i$ therefore take polynomial time and space
in the output blocklength $N_i$.

For $S$ ($h=2$), the same construction gives
$H_2^{-1}(1/2)$.

For $Z$ ($h=1$), the Hamming-ball estimate in
Eq.~\eqref{eq:si-Phi1} instead gives
$\mathsf R_0+H_2(\mathsf d_0)+\mathsf d_0\log_2 3<1$.
The balanced endpoint is the root $\beta_\star\in(0,1/2)$ of
$\beta_\star+H_2(\beta_\star)+\beta_\star\log_2 3=1$.
{Here, the dyadic construction instead selects the largest $t_i=\mu_i/q_i\in(0,1/2)$ satisfying}
\[
 q_i^{q_i}3^{\mu_i}<2^{q_i-\mu_i}\mu_i^{\mu_i}(q_i-\mu_i)^{q_i-\mu_i},
\]
which is exactly the strict balanced inequality.
{Then, $t_i\to\beta_\star$, and the same argument gives $\mathsf R_\infty=\beta_\star$ and $\mathsf d_\infty\ge\beta_\star$.}

Given $h$ and the selected parameters $(M_i,K_i,d_i)$,
Algorithm~\ref{alg:si-search} enumerates candidate matrix pairs
and checks the rejection conditions exactly.
The returned matrices $\mathsf B,A,F$ specify a fully addressable
code with $N_i=hM_i$, $K_i$ logical qubits, and $D_i\ge d_i$.

Lemma~\ref{lem:si-finite} proves termination and correctness.
There are at most $2^{h[M^2+M(M+1)/2]}$ candidate matrix pairs.
For each candidate, enumerating all $2^{2hM}$ Weyl pairs,
computing their orders and costs, and checking the rejection
conditions take $2^{O_h(M)}$ time, since each pair requires only
polynomial modular-arithmetic work.
The deterministic running time is therefore $2^{O_h(N^2)}$.
Processing one Weyl pair at a time uses polynomial working space.
A uniformly sampled admissible matrix pair triggers no rejection
with probability at least $1-\Phi_h$. At the selected lengths,
$\Phi_h<1/2$, so fewer than two admissible matrix pairs are needed
on average, giving expected time $2^{O_h(N)}$.
To sample $\mathsf B$ uniformly from the invertible matrices,
sample its entries independently and repeat if its binary reduction
is singular. Counting linearly independent columns
over $\F_2$ shows that a uniformly sampled ring matrix is invertible with
probability $\prod_{j=1}^M(1-2^{-j})>1/4$, and conditioning on
invertibility gives the uniform distribution. This sampling step has polynomial
expected cost. Synthesis of the resulting encoder
and computation of arbitrary logical phase powers are polynomial
in $N$ \referenceedit{by the implementation described in
Sec.~\ref{sec:si-encoding}}.

\begin{algorithm}[H]
\caption{Exact code construction in exponential time}
\label{alg:si-search}
\noindent\textbf{Input:} fixed $h$ and integers $M,K,d$ for which
$\Phi_h(M,K,d)<1$.\par
\smallskip
\noindent\textbf{Output:} matrices $\mathsf B,A,F$ specifying a fully addressable
$((hM,2^K,\ge d))_2$ code.\par
\begin{enumerate}
\item Set $r=M-K$. Enumerate matrices $\mathsf B$ and symmetric $A$ over
      $\rh$ in lexicographic order. Discard $\mathsf B$ unless $\mathsf B\bmod2$
      is invertible, and otherwise set $G=\mathsf B_{1:r,:}$ and compute
      $\mathsf B^{-1}$.
\item Enumerate nonzero Weyl pairs $(a,b)$ one at a time and
      determine their exact orders $2^s$. Retain only pairs of total
      cost below $d$. For $h\ge2$, at $s=1$ use
      $\wt(\alpha)+\wt(\beta)$ for
      $(a,b)=2^{h-1}(\alpha,\beta)$; at $s=2$ use the local costs
      \referenceedit{defined in the proof of Lemma~\ref{lem:si-order-four}};
      at $s\ge3$ use
      $\wt_{\mathrm s}(a,b)$, the number of affected $h$-qubit registers.
      For $h=1$, use binary symplectic weight.
\item For $h\ge2$ and $s=1$, skip a Weyl pair if
      $(a\mathsf B^{-1})_{r+1:M}\ne0$. For every remaining
      pair, reject the candidate matrices when
      $G(b-aA)^{\mathsf T}=0\pmod{2^h}$.
\item Return the first matrix pair for which no Weyl pair triggers
      rejection, using
      Eq.~\eqref{eq:si-encoding} and the addressing matrix in Eq.~\eqref{eq:si-F}.
\end{enumerate}
\end{algorithm}

\section{Rate-distance upper bounds from addressability}
\label{arxiv:si-contents-14}
\label{sec:si-converse}

Let $A_2(n,d)$ be the largest cardinality of a binary classical code of length $n$ and minimum distance at least $d$.
A code of cardinality $2^k$ has encoding rate $k/n$. Thus, the largest rate at these parameters is $n^{-1}\log_2 A_2(n,d)$.
The corresponding asymptotic encoding rate function at relative distance $t$ is
\begin{equation}
    R_2(t)=\limsup_{n\to\infty}\frac{1}{n}\log_2 A_2\!\left(n,\lceil tn\rceil\right),
    \qquad 0\le t\le\frac12.
    \label{eq:si-classical-rate}
\end{equation}

 Define
\begin{align}
 \mathcal H(x)&=H_2\!\left(\frac{1-\sqrt{1-x}}2\right),\notag\\
 \mathcal U(t)&=\min_{1-2t\le u\le1}
 \left\{\mathcal H(u^2)
       -\mathcal H\bigl(u(u-1+2t)\bigr)\right\},
 \quad 0<t<\frac12,
 \label{eq:si-MQC}
\end{align}
where $H_2(p)=-p\log_2 p-(1-p)\log_2(1-p)$ is the binary entropy function.
Corollary~23 of Alrabiah and Guruswami~\cite{Omar_Binary_2026}
gives the asymptotic upper bound $R_2(t)\le\calU(t)$ for $0<t<1/2$. Minimizing over $u$ in Eq. \eqref{eq:si-MQC} gives $\calU(t)$. The substitution $u=1-2t+2ts$, where $s\in[0,1]$, expresses the minimum over a fixed compact interval with a jointly continuous objective. Thus, $\calU$ extends continuously to $[0,1/2]$, with $\calU(0)=1$ and $\calU(1/2)=0$.

 {This bound also applies to arbitrary code families whose relative distances are specified by a lower limit.}
{Consider} binary classical codes with lengths $n_i\to\infty$, cardinalities $2^{k_i}$, and minimum distances $\Delta_i$.
{
If $\liminf_i\Delta_i/n_i\ge t$ for $0<t\le1/2$, then for every $0<\varepsilon<t$ we have $\Delta_i/n_i\ge t-\varepsilon$ for all sufficiently large $i$.
Consequently, $2^{k_i}\le A_2\!\left(n_i,\lceil(t-\varepsilon)n_i\rceil\right)$ for these indices.
Taking the upper limit gives $\limsup_i k_i/n_i\le R_2(t-\varepsilon)\le\calU(t-\varepsilon)$.
Letting $\varepsilon\downarrow0$ and using continuity yields
}
\begin{equation}
    \limsup_i\frac{k_i}{n_i}\le\calU(t)
    \quad\text{if}\quad
    \liminf_i\frac{\Delta_i}{n_i}\ge t,
    \qquad 0\le t\le\frac12.
    \label{eq:si-classical-family-bound}
\end{equation}
{For $t=0$, the same bound follows directly from $2^{k_i}\le2^{n_i}$ and $\calU(0)=1$.}

To apply this bound to fully addressable quantum codes, we
extract a classical binary code from the addressing matrix.
The following theorem shows that this code has dimension $K$
and minimum distance at least $D$, thereby relating the
classical rate bound to the quantum code parameters.

\begin{theorem}
\label{thm:si-upper}
Let  {$E$ be an encoding isometry for a} $((N,2^K,D))_2$ code {with} $K\ge1$.
{Suppose that $E$ is fully addressable at level} $h\ge1${, with addressing matrix $F$, in the sense of} Definition~\ref{def:si-address}.
{Let} $s_F$ {denote the number of} nonzero columns of $F\bmod2$.
{Then, deleting the zero columns of $F\bmod2$ yields a generator matrix for} a binary {linear} $[s_F,K,\ge D]$ code.
Consequently,
\begin{equation}
 2^K\le A_2(s_F,D),\qquad K+D-1\le s_F.
 \label{eq:si-finite-converse}
\end{equation}
\end{theorem}
\begin{proof}
Substitute $c=2^{h-1}v$, $v\in\F_2^K$, in~Eq.~\eqref{eq:si-address} to obtain $Z^{v(F\bmod2)}E=EZ^v$.
For $v\ne0$, {we have $Z^v\ne I$}, so $v(F\bmod2)\ne0$.
 {Thus, the rows of $F\bmod2$ are linearly independent, so its row rank is $K$.}
 {By the definition of distance $D$, every Pauli operator $P$ of weight less than $D$ satisfies $E^\dagger P E=\lambda_P I$ for some scalar $\lambda_P$.}
{Since $Z^v$ has eigenvalues $+1$ and $-1$ for $v\ne0$, the identity $E^\dagger Z^{v(F\bmod2)}E=Z^v$ implies} $\wt(v(F\bmod2))\ge D$.

Removing zero columns preserves these weights and the dimension, yielding a binary linear $[s_F,K,\ge D]$ code.
{Therefore, $2^K\le A_2(s_F,D)$ by the definition of $A_2$.}
The classical Singleton {bound}~\cite{Florence_The_1977} {gives $K+D-1\le s_F$}.
\end{proof}

Consider a fully addressable code family with $N_i\to\infty$
and $\mathsf R_\infty>0$. Pad the classical codes from
Theorem~\ref{thm:si-upper} with zero columns to length $N_i$.
This preserves their dimensions and minimum distances, giving
binary $[N_i,K_i,\ge D_i]$ codes.
The positive rate and the Plotkin bound
\cite[Corollary 19]{Omar_Binary_2026} imply $\mathsf d_\infty<1/2$.
Applying Eq.~\eqref{eq:si-classical-family-bound} with
$t=\mathsf d_\infty$ gives
\begin{equation}
 \mathsf R_\infty
 \le\limsup_i\frac{K_i}{N_i}
 \le\mathcal U(\mathsf d_\infty).
 \label{eq:si-universal-converse}
\end{equation}

The classical rate bound yields the following explicit upper bound
on the common coefficient $\eta$.

\begin{lemma}
\label{lem:si-ceiling}
{For each fixed $h\ge1$, consider a family of quantum codes with physical blocklengths $N_i\to\infty$.
Suppose that each code admits an encoding isometry that is fully addressable at level $h$ in the sense of Definition~\ref{def:si-address}.
Then, the common asymptotic coefficient $\eta=\min\{\mathsf R_\infty,\mathsf d_\infty\}$ satisfies}
\begin{equation}
 \eta\le c_{\mathrm{up}}=0.282652727463.
 \label{eq:si-ceiling}
\end{equation}
\end{lemma}
\begin{proof}
The case $\eta=0$ is immediate, so assume $\eta>0$.
Set $t_0=c_{\mathrm{up}}$ and $u_0=0.454058689$, so
$1-2t_0<u_0<1$. We first certify $\mathcal U(t_0)<t_0$.
Enclose each square root between adjacent multiples of $10^{-24}$
using integer square roots. Reduce logarithm arguments to $[1,2]$
and, with $z=(x-1)/(x+1)$, use
\[
 0\le\ln x-2\sum_{j=0}^{39}\frac{z^{2j+1}}{2j+1}
 \le\frac{2z^{81}}{81(1-z^2)}.
\]
The same enclosure at $x=2$ converts natural logarithms to base two.
Monotonicity of $H_2$ on $[0,1/2]$ propagates the square-root
intervals. Rational interval evaluation gives
\begin{equation}
 \begin{split}
 0.2826527274543
 &<\mathcal H(u_0^2)
   -\mathcal H\bigl(u_0(u_0-1+2t_0)\bigr)\\
 &<0.2826527274544<t_0.
 \end{split}
 \label{eq:si-MQC-witness}
\end{equation}

Since $u_0$ is admissible in the minimization in
Eq.~\eqref{eq:si-MQC}, the witness in
Eq.~\eqref{eq:si-MQC-witness} implies $\mathcal U(t_0)<t_0$.
Moreover, $\mathcal U$ is nonincreasing: increasing $t$ enlarges
the admissible interval in Eq.~\eqref{eq:si-MQC}, while the
objective at each previously admissible $u$ does not increase,
since $\mathcal H$ is increasing.
If $\eta\ge t_0$, both $\mathsf R_\infty$ and
$\mathsf d_\infty$ would be at least $t_0$, so
Eq.~\eqref{eq:si-universal-converse} would imply
\[
 t_0\le\mathsf R_\infty
 \le\mathcal U(\mathsf d_\infty)
 \le\mathcal U(t_0)<t_0,
\]
a contradiction. Hence $\eta\le c_{\mathrm{up}}$.
\end{proof}

\referenceedit{The addressing form in Eq.~\eqref{eq:si-F} strengthens
this bound by shortening the associated classical code.}

\begin{corollary}
\label{cor:si-digit-ceiling}
For a family as in Lemma~\ref{lem:si-ceiling}, suppose $N_i=hM_i$
and $(F_i)_{\ell,(j,b)}=2^bz_{\ell,j}^{(i)}\pmod{2^h}$ for
$0\le b<h$. Then
\begin{equation}
 \eta\le\frac{c_{\mathrm{up}}}{h}.
 \label{eq:si-digit-ceiling}
\end{equation}
\end{corollary}
\begin{proof}
Again, assume $\eta>0$, since otherwise the claim is immediate.
All columns with $b\ge1$ vanish modulo two, so each register
contributes at most one nonzero column and $s_{F_i}\le M_i$.
Theorem~\ref{thm:si-upper} therefore gives a classical
$[M_i,K_i,\ge D_i]$ code after padding with zeros.
Its lower-limit rate is $h\mathsf R_\infty$, and its lower-limit
relative distance is at least $h\mathsf d_\infty$, since $N_i=hM_i$.
The positive rate and the Plotkin bound
\referenceedit{\cite[Corollary 19]{Omar_Binary_2026}} ensure
$h\mathsf d_\infty<1/2$.
By the definition of $A_2$ and
Eq.~\eqref{eq:si-classical-family-bound}, applied with
$n_i=M_i$ and $t=h\mathsf d_\infty$,
\begin{equation}
 2^{K_i}\le A_2(M_i,D_i),\qquad
 h\mathsf R_\infty\le\mathcal U(h\mathsf d_\infty).
 \label{eq:si-digit-converse}
\end{equation}
The proof of Lemma~\ref{lem:si-ceiling} establishes that
$\mathcal U$ is nonincreasing and
$\mathcal U(c_{\mathrm{up}})<c_{\mathrm{up}}$.
If $h\eta\ge c_{\mathrm{up}}$, Eq.~\eqref{eq:si-digit-converse}
would therefore give
\[
 c_{\mathrm{up}}\le h\mathsf R_\infty
 \le\mathcal U(h\mathsf d_\infty)
 \le\mathcal U(c_{\mathrm{up}})<c_{\mathrm{up}},
\]
a contradiction. Hence $\eta\le c_{\mathrm{up}}/h$.
\end{proof}

\section{Quantitative comparison with prior constructions}
\label{arxiv:si-contents-15}
\label{sec:si-prior}

We compare fully addressable transversal $T$ gates implemented by
physical $T$ powers on one fixed encoding. The parameters below are
the asymptotic encoding rate $\mathsf R_\infty$ and relative distance
$\mathsf d_\infty$, with blocklength measured in physical qubits.

San-Jos\'e constructs binary CSS codes by converting higher-order
Schur-product conditions on AG codes into weighted binary divisibility
\cite[Proposition 4.5 and Theorem 6.4]{Rodrigo_Asymptotically_2026}.
For addressable $T$, we apply the hierarchy-lowering argument of
\referenceedit{Ref.~\cite[Remark 6.8]{Rodrigo_Asymptotically_2026}}
to his weighted level-four construction. If $w_p$ are its
physical phase weights, $g_a$ its logical $X$ rows, and
$\epsilon_a\in\{1,-1\}$ its logical phase signs, the addressing
coefficients are $F_{a,p}=\epsilon_aw_pg_{a,p}\pmod8$.
This weighted form preserves the original blocklength.
\referenceedit{We specialize the level-four construction to $m=18$ and
$\tau=128$. Proposition~4.5 of
Ref.~\cite{Rodrigo_Asymptotically_2026}, applied with $j=8$,
bounds the binary conversion length by
$\sum_{a=1}^{8}\binom{18}{a}=106761$.
Choosing the logical-coordinate fraction $\lambda=257/523264$
in that reference's Eqs.~(28)--(29) then gives}
\begin{equation}
 \mathsf R_\infty\ge\frac{257}{55836750327},\qquad
 \mathsf d_\infty\ge\frac{257}{55672124865}.
 \label{eq:si-prior-T-guarantee}
\end{equation}
This explicit substitution therefore guarantees both quantities
to be at least $4.6027033897\ldots\times10^{-9}$.

Lin, Wu, and Cheng combine AG codes with a binary embedding to
obtain generalized divisibility, which directly gives fully
addressable transversal $T$ powers on CSS codes
\cite[Theorem 4.5]{Tongyin_Asymptotically_2026}.
Their explicit construction uses a binary $[244,10,100]$ embedding
and has $K_i=\lfloor n_i/992\rfloor$ and
$N_i=244(n_i-K_i)$, where $n_i$ is the outer AG code length.
\referenceedit{Corollary~D.1 of
Ref.~\cite{Tongyin_Asymptotically_2026}} gives
\begin{equation}
 \mathsf R_\infty=\frac1{241804},\qquad
 \mathsf d_\infty\ge\frac1{241804}.
 \label{eq:si-prior-LWC}
\end{equation}
The common guarantee is $\simeq 4.1355808837\times10^{-6}$.

Meiburg obtains fully addressable transversal $T$ gates on CSS
codes by reserving AG evaluation points for logical labels and
using interpolation and a binary embedding to control their phases
\cite[Secs.~3.3--3.5]{Alexander_Constant_2026}.
\referenceedit{Before the coordinate-repetition step, the four-bit
packing in Ref.~\cite[Sec.~3.5]{Alexander_Constant_2026} gives}
$K_i=4t_i$ and $N_i=385(n_i-t_i)$, with $t_i/n_i\to5/186$,
where $n_i$ is the total number of AG evaluation points and $t_i$
counts the reserved points. These parameters already implement
arbitrary logical $T$ powers using physical $T$ powers, and yield
\begin{equation}
 \mathsf R_\infty=\frac4{13937},\qquad
 \mathsf d_\infty\ge\frac1{13937}.
 \label{eq:si-prior-Meiburg}
\end{equation}
Thus the common guarantee is $1/13937\simeq7.1751\times10^{-5}$.

Meiburg's modular-clock construction also applies to $T$~\cite[Sec.~3.6]{Alexander_Constant_2026}.
\referenceedit{We specialize it to a Garcia--Stichtenoth tower over
$\F_{1024}$, for which $g_i/n_i\to1/31$
\cite{Arnaldo_On_1996}, using Meiburg's stated divisor choices
$a_{0,i}/n_i\to1/3$ and $a_{1,i}/n_i\to2/3$.}
Binary basis expansion gives $10n_i$ clock coordinates and $10(a_{1,i}-a_{0,i})$ logical clocks.
Lifting to $\Z/8\Z$ preserves the distance lower bound $\min\{n_i-a_{1,i},a_{0,i}-2g_i+2\}$.
\referenceedit{The local encoding in
Ref.~\cite[Eq.~(18)]{Alexander_Constant_2026} uses
$(\lceil\log_2 8\rceil+1)(8-1)=28$ physical qubits per clock
and preserves this distance bound.}
Restricting each logical clock to its two basis states labelled zero and one gives $N_i=280n_i$ and $K_i=10(a_{1,i}-a_{0,i})$ without decreasing the distance.
Thus,
\begin{equation}
    \mathsf R_\infty=\frac1{84},\qquad
    \mathsf d_\infty\ge\frac{1/3-2/31}{280}=\frac5{5208}.
    \label{eq:si-prior-Meiburg-clock}
\end{equation}
The common guarantee is $5/5208\simeq9.6006\times10^{-4}$.

All four prior constructions have positive asymptotic rate and relative distance.
The prior-work entries in the comparison table (Sec.~I of the main text) use the smaller of the two guarantees for each stated parameter choice.
The prior constructions and our constructions meet the same logical and physical $T$-gate requirement.
Our polynomial-time and exponential-time constructions
\referenceedit{in Secs.~\ref{sec:si-ag-parameters} and~\ref{sec:si-exact-parameters}}
guarantee both quantities at least $52/2025$ and $t_\star/3$, respectively.

\section{Finite examples and their exact certificates}
\label{arxiv:si-contents-16}
\label{sec:si-examples}

The codes $\mathcal Q_{48}=((48,16,5))_2$ and
$\mathcal Q_{90}=((90,256,7))_2$ use~Eq.~\eqref{eq:si-encoding} with
$h=3$ and $(M,r,K)=(16,12,4)$ and $(30,22,8)$, respectively.
In each case $\mathsf B$ is invertible over $\mathbb Z/8\mathbb Z$, $A$ is
symmetric, and the quadratic exponent is evaluated modulo sixteen.
The matrices were obtained by bounded candidate sampling
\referenceedit{using the rejection tests of Algorithm~\ref{alg:si-search}}.
Checking every nonzero Weyl pair of cost below the respective
claimed distance triggers no rejection, certifying both distance
lower bounds \referenceedit{by the proof of Lemma~\ref{lem:si-finite}}.
The $90$-qubit target also satisfies the sufficient
ensemble condition $\Phi_3<1$, \referenceedit{where $\Phi_3$ is
defined in Eq.~\eqref{eq:si-Phi}}.

\subsection{Addressable transversal \texorpdfstring{$T$}{T} gates}
\label{arxiv:si-contents-17}
\label{sec:si-example-specifications}

The $j$th three-qubit register contains physical qubits $(3j-2,3j-1,3j)$ in
least-to-most-significant digit order. With $G,L$ the first $r$
and last $K$ rows of $\mathsf B$, \referenceedit{collect the selectors
of Lemma~\ref{lem:si-encoding} in}
\begin{equation}
 Z_{\mathrm{addr}}=(\mathsf B^{-\mathsf T})_{r+1:M,:},\qquad
 GZ_{\mathrm{addr}}^{\mathsf T}=0,\quad
 LZ_{\mathrm{addr}}^{\mathsf T}=I_K\pmod8.
 \label{eq:si-finite-selectors}
\end{equation}
For $c\in\mathcal R_3^K$, put $a=cZ_{\mathrm{addr}}\pmod8$.
The physical layer \referenceedit{specified by Eq.~\eqref{eq:si-F} at $h=3$}
\begin{equation}
 \bigotimes_{j=1}^M
 (T^{a_j}\otimes T^{2a_j}\otimes T^{4a_j})
 \label{eq:si-finite-phase-layer}
\end{equation}
has phase $e^{2\pi i cx^{\mathsf T}/8}$ on every term of
$E\ket x$, so it implements $\bigotimes_\ell T^{c_\ell}$ exactly.
Combining exponents modulo eight gives at most one single-qubit
phase gate on each physical qubit.

\subsection{Verification of code distances}
\label{arxiv:si-contents-18}
\label{sec:si-finite-distance}
\label{sec:si-minimum-witnesses}

We certify each distance by verifying detection of all Pauli errors
of weight below $d$ and exhibiting a weight-$d$ Pauli with
nonscalar logical action.

For $W(a,b)=\bigotimes_jX_8(a_j)Z_8(b_j)$,
\referenceedit{Eq.~\eqref{eq:si-phase-conjugation} gives}
\begin{equation}
 Q_A^\dagger W(a,b)Q_A
 =e^{-\pi i aAa^{\mathsf T}/8}W(a,b-aA),
 \label{eq:si-finite-conjugation}
\end{equation}
\referenceedit{and Lemma~\ref{lem:si-compression} requires
$G(b-aA)^{\mathsf T}=0\pmod8$ for nonzero compression.}

\referenceedit{At exact order two, write $(a,b)=4(\alpha,\beta)$ with
$\alpha,\beta\in\F_2^M$ not both zero.
By Lemma~\ref{lem:si-compression}, nonzero compression requires
$(a\mathsf B^{-1})_{r+1:M}=x'-x$ for some binary logical labels
$x,x'$. The left-hand side has entries in $\{0,4\}$, whereas
the right-hand side has entries in $\{0,1,7\}$ modulo eight.
Their common value must therefore be zero. The order-two
rejection conditions of Sec.~\ref{sec:si-compression-proof}
are consequently
\[
 G(b-aA)^{\mathsf T}=0,\qquad
 (a\mathsf B^{-1})_{r+1:M}=0\pmod8.
\]

Put $G_2=G\bmod2$ and $A_2=A\bmod2$.
Using $4z=0\pmod8$ if and only if $z=0\pmod2$, the first
condition becomes
\[
 G_2(\beta-\alpha A_2)^{\mathsf T}
 =G_2\beta^{\mathsf T}+G_2A_2\alpha^{\mathsf T}=0
\pmod 2,
\]
where we used $A_2^{\mathsf T}=A_2$ and the equality of addition
and subtraction in $\F_2$.
Reducing and transposing the second condition similarly gives
$J\alpha^{\mathsf T}=0$, where
$J=((\mathsf B^{-1})_{:,r+1:M})^{\mathsf T}\bmod2$.
Combining these two binary equations gives
\begin{equation}
 \begin{pmatrix}G_2A_2&G_2\\J&0\end{pmatrix}
 \begin{pmatrix}\alpha^{\mathsf T}\\\beta^{\mathsf T}\end{pmatrix}=0 \pmod2
 \label{eq:si-parity-certificate}
\end{equation}
By the proof of Lemma~\ref{lem:si-order-two}, the physical Pauli weight is
$\wt(\alpha)+\wt(\beta)$. The order-two test therefore rejects
a candidate exactly when Eq.~\eqref{eq:si-parity-certificate}
has a nonzero solution with $\wt(\alpha)+\wt(\beta)<d$.

For $\mathcal Q_{48}$ and $\mathcal Q_{90}$, exact verification
confirms that Eq.~\eqref{eq:si-parity-certificate} has no
nonzero solution with $\wt(\alpha)+\wt(\beta)<d$,
for $d=5$ and $d=7$, respectively.}

\referenceedit{For exact order four, write $(a,b)=2(a',b')$ with
$a',b'\in(\Z/4\Z)^M$ and at least one odd coordinate.
The rejection condition in Sec.~\ref{sec:si-compression-proof} becomes
\[
 G(b'-a'A)^{\mathsf T}=0\pmod4.
\]
As in the proof of Lemma~\ref{lem:si-order-four}, assign each
local pair weight zero at $(0,0)$, one on $\mathcal A_3$ from
Eq.~\eqref{eq:si-localsets}, and two otherwise; retain only
pairs whose total assigned weight is below $d$.

At exact order eight, $(a,b)$ has at least one odd coordinate.
Combining the same rejection condition with the weight
restriction of Lemma~\ref{lem:si-higher-order} gives
\[
 G(b-aA)^{\mathsf T}=0\pmod8,\qquad
 |\{j:(a_j,b_j)\ne(0,0)\}|<d.
\]
For both codes, exact verification excludes all solutions
under these order and weight restrictions, with $d=5$ and
$d=7$, respectively.

Together with the order-two result, these exclusions imply
$E^\dagger PE=0$ for every nonidentity qubit Pauli $P$ of
weight below $d$, by the proof of Lemma~\ref{lem:si-finite}.
Hence $D_{48}\ge5$ and $D_{90}\ge7$.}

Matching upper bounds follow from a common criterion: a Pauli
$P_N$ with nonscalar compression $E_N^\dagger P_NE_N$ implies
$D_N\le\wt(P_N)$. For the two codes, take
\begin{align}
 P_{48}&=X_9X_{24}X_{48}Z_{34}Z_{40},
 \label{eq:si-q48-witness}
 \\
 P_{90}&=X_{24}X_{66}X_{69}X_{72}Z_{25}Z_{46}Z_{76}.
 \label{eq:si-q90-witness}
\end{align}
For the archived matrices, Eq.~\eqref{eq:si-compression-value} gives
\begin{equation}
 P_NE_N=E_NZ_2Z_4,\qquad N\in\{48,90\},
 \label{eq:si-witness-actions}
\end{equation}
where $Z_2Z_4$ acts on the logical qubits and is nonscalar.
The witness weights are five and seven, respectively, so
$D_{48}\le5$ and $D_{90}\le7$. Together with the lower bounds
above, this proves $D_{48}=5$ and $D_{90}=7$.

\subsection{Ancilla-free encoding circuit}
\label{arxiv:si-contents-19}
\label{sec:si-q48-circuit}

For $\mathcal Q_{48}$, place the four logical inputs on qubits
$37,40,43,46$ and initialize the other qubits to zero. Applying
$H$ to the first $36$ qubits prepares the uniform ring inputs.
The modular map $y\mapsto y\mathsf B$, followed by
\referenceedit{the quadratic phase $Q_A$ defined in the proof of
Lemma~\ref{lem:si-encoding}}, then implements
Eq.~\eqref{eq:si-encoding} using the $48$ data qubits.

The linear map acts on $16$ three-qubit registers. Gauss--Jordan
elimination of $\mathsf B$ over $\mathcal R_3$ gives a factorization
implemented by $8$ register swaps, $12$ multiplications
$x\mapsto ux\bmod8$ with odd $u$, and $202$ weighted additions
$(x,y)\mapsto(x,y+cx\bmod8)$.
Each register swap consists of three two-qubit SWAP gates,
giving the $24$ SWAP gates reported in Sec.~V of the main text.

Expanding $Q_A$ in the binary digits of the registers gives
single-qubit phase gates and $466$ two-qubit controlled-phase
gates.

\begingroup
\small

\endgroup
\end{document}